\documentclass[aps,prd,preprintnumbers,superscriptaddress,nofootinbib]{revtex4}
\usepackage{dcolumn}
\usepackage{amsmath}
\usepackage{amssymb}
\usepackage{amsthm}
\usepackage{mathtools}
\usepackage{mathrsfs}
\usepackage{bm}
\usepackage{slashed} 
\usepackage{graphicx}
\usepackage{multirow}
\usepackage{tikz}
\usepackage[caption=false]{subfig}
\usepackage{relsize}	
\usepackage{array}
\usepackage{float}
\usepackage{color}
\usepackage{xcolor}
\usepackage{soul}
\usepackage{verbatim} 
\usepackage{siunitx}
\usepackage{hyperref}
\hypersetup{colorlinks=true,linkcolor=purple,anchorcolor=blue,citecolor=blue, filecolor=blue,urlcolor=red,bookmarksnumbered=true,
pdfview=FitB
}
\allowdisplaybreaks
\def\be{\begin{eqnarray}}
\def\ee{\end{eqnarray}}

\newcommand{\ud}{\text{d}}

\colorlet{purple1}{blue!70!red}
\colorlet{darkred}{red!50!black}

\usepackage{braket}

\begin{document}

\title{Angular momentum and generalized parton distributions for the proton with basis light-front quantization}

\author{Yiping~Liu}
\email{liuyipimg21@mails.ucas.ac.cn} 
\affiliation{Institute of Modern Physics, Chinese Academy of Sciences, Lanzhou 730000, China}
\affiliation{School of Nuclear Science and Technology, University of Chinese Academy of Sciences, Beijing 100049, China}

\author{Siqi~Xu}
\email{xsq234@impcas.ac.cn} \affiliation{Institute of Modern Physics, Chinese Academy of Sciences, Lanzhou 730000, China}
\affiliation{School of Nuclear Science and Technology, University of Chinese Academy of Sciences, Beijing 100049, China}

\author{Chandan~Mondal}
\email{mondal@impcas.ac.cn} 
\affiliation{Institute of Modern Physics, Chinese Academy of Sciences, Lanzhou 730000, China}
\affiliation{School of Nuclear Science and Technology, University of Chinese Academy of Sciences, Beijing 100049, China}

\author{Xingbo~Zhao}\email{xbzhao@impcas.ac.cn}
\affiliation{Institute of Modern Physics, Chinese Academy of Sciences, Lanzhou 730000, China}
\affiliation{School of Nuclear Science and Technology, University of Chinese Academy of Sciences, Beijing 100049, China}

\author{James~P.~Vary}\email{jvary@iastate.edu}
\affiliation{Department of Physics and Astronomy, Iowa State University,
Ames, IA 50011, U.S.A.}

\collaboration{BLFQ Collaboration}
\date{\today}

\begin{abstract}

We study the unpolarized and the helicity dependent generalized parton distributions (GPDs) for the valence quarks of the proton in both momentum space and position space within the basis light-front quantization (BLFQ) framework. The GPDs for the valence quarks are computed from the eigenvectors of a light-front effective Hamiltonian  in the valence Fock sector consisting of a three-dimensional confinement potential and a one-gluon exchange interaction with fixed coupling. Employing these GPDs,  we obtain  the spatial distributions of quark angular momentum inside the proton. In our BLFQ approach, we explore various definitions of angular momentum density and illustrate the differences between them arising from terms that integrate to zero. We also discuss the  flavor contributions to the quark angular momentum densities.

\end{abstract}
\maketitle
\section{Introduction}
The origin of nucleon spin is one of the major puzzles in
modern particle physics. The  well known European Muon Collaboration experiment~\cite{EuropeanMuon:1987isl} has triggered interest in understanding the nucleon spin from the contributions of the spin and the orbital angular momentum (OAM) of each of its constituents. 
 In this context, how the total angular momentum (TAM) is split into separate quark and gluon (partons) contributions is intrinsically debatable due to quark-gluon couplings and the non-uniqueness of the decomposition~\cite{Leader:2013jra,Wakamatsu:2014zza,Liu:2015xha}. Meanwhile, it has become clear that the generalized parton distributions (GPDs)~\cite{Diehl:2003ny,Belitsky:2005qn,Goeke:2001tz}, appearing in the description of hard exclusive reactions like deeply virtual Compton scattering or deeply virtual meson production, provide us with essential information about the spatial distributions and orbital motion of partons inside the nucleon, and allow us to draw three-dimensional pictures of the nucleon. For more than two decades, the GPDs have been attracting numerous dedicated experimental and theoretical efforts as many observables can be connected to them. The GPDs are functions of three variables, namely, longitudinal momentum fraction ($x$) of the constituent, the skewness ($\zeta$) or the longitudinal momentum transferred, and the square of the total momentum transferred ($t$). Their first moments are linked to the electromagnetic form factors, whereas they reduce to the ordinary parton distributions in the forward limit ($t=0$). The second moments of the GPDs correspond to the gravitational form factors, which are linked to matrix elements of the energy-momentum tensor (EMT). Being off-forward matrix elements, the GPDs do not have probabilistic interpretations. Meanwhile, for zero skewness the Fourier transform (FT) of the GPDs with respect to the momentum transfer in the transverse direction provides the impact parameter dependent GPDs that do have a   probabilistic interpretation~\cite{Burkardt:2000za,Burkardt:2002hr}. 
 The impact parameter dependent GPDs encode the correlations in
spatial and momentum distributions of partons in the nucleon. They contain the information about partonic distributions in the transverse position space for a given longitudinal momentum fraction carried by the constituent. 

Ji has shown that the partonic contribution to the total angular momentum of the nucleon can be calculated using the second moment of the GPDs~\cite{Ji:PRL}. 
Since, the GPDs provide the spatial distribution of the constituents inside the nucleon, it is therefore credible that the GPDs carry also the knowledge about the spatial distribution of angular momentum~\cite{Lorce:2017wkb,Polyakov:2002yz,Adhikari:2016dir,Kumar:2017dbf}. The angular momentum distribution in three-dimensional coordinate space was first introduced in Ref.~\cite{Polyakov:2002yz}. However, there is an issue of relativistic corrections for the three-dimensional distribution, while this ambiguity can be avoided by defining the two-dimensional distribution in the infinite momentum frame~\cite{Adhikari:2016dir,Leader:2013jra}. Different techniques to calculate the angular momentum distributions in the transverse plane have been prescribed in Ref.~\cite{Adhikari:2016dir} and concluded that none of them agrees at the density level. Meanwhile, a more detailed discussion on the various definitions of angular momentum has been reported in Ref.~\cite{Lorce:2017wkb} and the authors have identified all the missing terms, which hinder the proper comparison. They have illustrated explicitly using a scalar diquark model that there is no discrepancy between the different definitions of angular momentum densities. Later, the distributions of quark angular momentum in a light-front quark-diquark model (with both scalar and axial vector diquark) motivated by soft wall AdS/QCD have been investigated in Ref.~\cite{Kumar:2017dbf}.

In this paper, we investigate the spatial distributions of quark angular momentum inside the proton from its valence light-front wave functions (LFWFs) that features all three active quarks’ spin, flavor, and three-dimensional spatial information on the
same footing.  Our theoretical framework to explore the nucleon structure is rooted in  basis light front quantization (BLFQ)~\cite{Vary:2009gt}, which provides a computational framework for solving
relativistic many-body bound state problem in quantum field theories~\cite{Vary:2009gt,Li:2021jqb,Zhao:2014xaa,Wiecki:2014ola,Li:2015zda,Li:2017mlw,Jia:2018ary,Lan:2019vui,Lan:2019rba,Lekha,Tang:2018myz,Tang:2019gvn,Mondal:2019jdg,Xu:2021wwj,Lan:2019img,Qian:2020utg,Lan:2021wok}. We evaluate the valence quark GPDs of the proton in both momentum space and position space using the LFWFs based on the BLFQ with only the valence Fock sector of the proton considered. The BLFQ provides for a Hamiltonian formalism that incorporates the advantages of the light-front dynamics~\cite{Brodsky:1997de}. Our effective Hamiltonian includes a three-dimensional confinement potential consisting of the light-front holography in the transverse direction~\cite{Brodsky:2014yha}, a longitudinal confinement~\cite{Li:2015zda}, and a one-gluon exchange interaction with fixed coupling to account for the spin structure~\cite{Mondal:2019jdg}. The nonperturbative solutions for the three-body LFWFs are given by the recent BLFQ study of the nucleon~\cite{Mondal:2019jdg}. These  LFWFs have been applied successfully to  predict the electromagnetic and axial form factors, radii, parton  distribution functions (PDFs), and many other quantities of the nucleon~\cite{Mondal:2019jdg,Xu:2021wwj,Mondal:2021wfq}. Here, we extend those investigations to study the proton GPDs and their application for the description of angular momentum distributions.

The paper is organized as follows.  We briefly summarize the BLFQ formalism for the nucleon in Sec.~\ref{sec:formalism}. We then present a  detailed description of the angular momentum and the associated GPDs in Sec.~\ref{sec:AM}. Sec.~\ref{sec:results} details our numerical results for the GPDs, and different angular momentum densities. At the end, we provide a brief summary and conclusions in Sec.~\ref{sec:summary}.
\section{Light-front effective Hamiltonian for the proton}\label{sec:formalism}
The LFWFs that encode the structure of hadronic bound states are obtained as the eigenfunctions of the eigenvalue equation of the Hamiltonian: 
$
H_{\rm LF}\vert \Psi\rangle=M_{\rm h}^2\vert \Psi\rangle,
$
where $H_{\rm LF}$ represents the light-front Hamiltonian of the hadron with the mass squared ($M_{\rm h}^2$) eigenvalue. With quarks being the only explicit degree of freedom, the effective Hamiltonian we employ for the proton includes the two-dimensional harmonic oscillator (`2D-HO') transverse confining potential along with a longitudinal confinement and an effective one-gluon exchange interaction~\cite{Mondal:2019jdg}
\begin{align}\label{hami}
H_{\rm eff}=&\sum_a \frac{{\vec k}_{\perp a}^2+m_{a}^2}{x_a}+\frac{1}{2}\sum_{a\ne b}\kappa^4 \Big[x_ax_b({ \vec r}_{\perp a}-{ \vec r}_{\perp b})^2-\frac{\partial_{x_a}(x_a x_b\partial_{x_b})}{(m_{a}+m_{b})^2}\Big]
\nonumber\\&+\frac{1}{2}\sum_{a\ne b} \frac{C_F 4\pi \alpha_s}{Q^2_{ab}} \bar{u}(k'_a,s'_a)\gamma^\mu{u}(k_a,s_a)\bar{u}(k'_b,s'_b)\gamma^\nu{u}(k_b,s_b)g_{\mu\nu}\,,
\end{align}
where $x_a$ and ${\vec k}_{\perp a}$ represent the longitudinal momentum fraction and the relative transverse momentum carried by quark $a$. $m_{a}$ is the mass of the quark a, and $\kappa$ defines the strength of the confinement. The variable $\vec{r}_\perp={ \vec r}_{\perp a}-{ \vec r}_{\perp b}$ is the transverse separation between two quarks. The last term in the effective Hamiltonian corresponds to the OGE interaction where $Q^2_{ab}=-q^2=-(1/2)(k'_a-k_a)^2-(1/2)(k'_b-k_b)^2$ is the average momentum transfer squared, $C_F =-2/3$ is the color
factor, $\alpha_s$ is the coupling constant and $g_{\mu\nu}$ is the metric tensor. ${u}(k_a,s_a)$ represents the spinor with momentum $k_a$ and spin $s_a$.

For the BLFQ basis representation, the 2D-HO function is adopted for the transverse direction, while we employ the discretized plane-wave basis in the longitudinal direction~\cite{Vary:2009gt,Zhao:2014xaa}. Diagonalizing the Hamiltonian, Eq.~(\ref{hami}), in our chosen basis space gives the eigenvalues as squares of the bound state eigenmasses, and the eigenstates which specify the LFWFs. The lowest eigenstate is naturally identified as the nucleon state, denoted as $\ket{P, {\Lambda}}$, with $P$ and $\Lambda$ being the momentum and the helicity of the state. 
In terms of the basis function the LFWFs of the nucleon are expressed as 
\begin{align}
\Psi^{\Lambda}_{\{x_i,\vec{k}_{i\perp},\lambda_i\}}=\sum_{\{n_i,m_i\}} \psi^{\Lambda}_{\{x_{i},n_{i},m_{i},\lambda_i\}} \prod_i \phi_{n_i,m_i}(\vec{k}_{i\perp};b) \,,\label{wavefunctions}
\end{align}
where $\psi^{\Lambda}_{\{x_{i},n_{i},m_{i},\lambda_i\}}=\braket{P, {\Lambda}|\{x_i,n_i,m_i,\lambda_i\}}$ is the LFWF in the BLFQ basis obtained by diagnalizing Eq.~(\ref{hami}) numerically. The  2D-HO function we adopt as the transverse basis function is
\begin{align}
\phi_{n,m}(\vec{k}_{\perp};b)
 =\frac{\sqrt{2}}{b(2\pi)^{\frac{3}{2}}}\sqrt{\frac{n!}{(n+|m|)!}}e^{-\vec{k}_{\perp}^2/(2b^2)}\left(\frac{|\vec{k}_{\perp}|}{b}\right)^{|m|}L^{|m|}_{n}(\frac{\vec{k}_{\perp}^2}{b^2})e^{im\theta}\label{ho}\,,
\end{align}
with $b$ as its scale parameter; $n$ and $m$ are the principal and orbital quantum
numbers, respectively, and $L^{|m|}_{n}$ is the associated Laguerre polynomial. In the discretized plane-wave basis, the longitudinal momentum fraction $x$ is defined as
$
x_i=p_i^+/P^+=k_i/K,
$
where the dimensionless quantity signifying the choice of antiperiodic boundary conditions is $k=\frac{1}{2}, \frac{3}{2}, \frac{5}{2}, ...$ and $K=\sum_i k_i$. The multi-body basis states have selected values of the total angular momentum projection
$
M_J=\sum_i\left(m_i+\lambda_i\right),
$
where $\lambda$ is used to label the quark helicity. The transverse basis truncation is specified by the dimensionless parameters $N_{\rm max}$, such that $\sum_i (2n_i+| m_i |+1)\le N_{\rm max}$. The basis cutoff $N_{\rm max}$ acts implicitly as the ultraviolet (UV) and infrared (IR) regulators for the LFWFs in the transverse direction, with a UV cutoff $\Lambda_{\rm UV}\approx b \sqrt{N_{\rm max}}$ and an IR cutoff $\Lambda_{\rm IR}\approx b /\sqrt{N_{\rm max}}$. The longitudinal basis cutoff $K$ controls the numerical resolution and regulates the longitudinal direction.

Parameters in the model Hamiltonian are fixed to reproduce the ground state mass of the nucleon and to fit the Dirac flavor form factors~\cite{Xu:2021wwj}. The LFWFs in this model have been successfully applied to compute a wide class of different
and related nucleon observables, e.g., the electromagnetic and axial form factors, radii, PDFs, helicity asymmetries, transverse momentum dependent parton distribution functions etc., with remarkable overall success~\cite{Mondal:2019jdg,Xu:2021wwj,Mondal:2021wfq}.
\section{Angular momentum distributions} \label{sec:AM}
In this section, we introduce our notation and briefly review the derivation of angular momentum distribution following Ref.~\cite{Lorce:2017wkb}. In field theory, the generalized angular momentum tensor operator is written as follows
\begin{equation}
J^{\mu\alpha\beta}(y)=L^{\mu\alpha\beta}(y)+S^{\mu\alpha\beta}(y) \; . \label{J}
\end{equation}
Both of the  contributions are antisymmetric under $\alpha\leftrightarrow\beta$. When $\alpha,\beta$ are spatial components, $L^{\mu\alpha\beta}(y)$ and  $S^{\mu\alpha\beta}(y)$ are identified with the OAM and spin operators, respectively. The first contribution can be expressed in terms of the  EMT as
\begin{equation}
L^{\mu\alpha\beta}(y)= y^{\alpha}T^{\mu\beta}(y)-y^{\beta}T^{\mu\alpha}(y) \; .
\end{equation}
Note that $T^{\mu\nu}$ is referred to the canonical EMT and it is, in general, neither gauge invariant nor symmetric. Meanwhile, the TAM can also be expressed in a pure orbital form,
\begin{equation}
J^{\mu\alpha\beta}_{\text{Bel}}(y)=y^{\alpha}T_{\text{Bel}}^{\mu\beta}(y)-y^{\beta}T_{\text{Bel}}^{\mu\alpha}(y)\,,
\end{equation}
using the Belinfante-improved EMT~\cite{Belinfante1, Belinfante2,Rosenfeld}, which is defined by adding a term to the definition of $T^{\mu\nu}$ as
\begin{align}
T^{\mu\nu}_{\text{Bel}}(y)&= T^{\mu\nu}(y)+\partial_{\lambda}G^{\lambda\mu\nu}(y)\,,\label{tbel}
\end{align}
where  $G^{\lambda\mu\nu}$ is given by
\begin{equation}\label{G}
G^{\lambda\mu\nu}(y)=\frac{1}{2}\left[S^{\lambda\mu\nu}(y)+S^{\mu\nu\lambda}(y)+S^{\nu\mu\lambda}(y)\right]=-G^{\mu\lambda\nu}(y) \;.
\end{equation}
The additional term revises the definition of the local density without changing the TAM. The Belinfante-improved tensor $T^{\mu\nu}_{\text{Bel}}$ is conserved, symmetric and gauge invariant. 
The Belinfante-improved tensors can be seen as effective densities, where the effects of spin are imitated by a superpotential contribution to the angular momentum. In order to determine the angular momentum distributions, we are interested in the matrix elements of the above mentioned operator densities.

For a spin-$\frac{1}{2}$ target, the matrix elements of the general local asymmetric $T^{\mu\nu}$  are parametrized in terms of several gravitational form factors~\cite{Leader:2013jra}:
\begin{align}
&\langle P', {\bf\Lambda}'\lvert T^{\mu\nu}(0) \rvert P, {\bf\Lambda}\rangle  =  \bar{u}(P', {\bf\Lambda}')\Big[\frac{\bar{P}^{\mu}\bar{P}^\nu}{M}\,A(t)+\frac{\bar{P}^{\mu}i\sigma^{\nu\lambda}\Delta_{\lambda}}{4M}\,(A+B+D)(t)\nonumber\\ &\quad\quad\quad\quad
+\frac{\Delta^{\mu}\Delta^{\nu}-g^{\mu\nu}\Delta^2}{M}\,C(t)+Mg^{\mu\nu}\,\bar{C}(t)+\frac{\bar{P}^{\nu}i\sigma^{\mu\lambda}\Delta_{\lambda}}{4M}\,(A+B-D)(t)\Big]u(P, {\bf\Lambda}) \; , \label{dec4}
\end{align} 
where $\bar{P}=\frac{1}{2}(P'+P)$, $\Delta=P'-P$, $t=\Delta^2$,  $M$ is the system mass, the three-vector ${\bf\Lambda}({\bf\Lambda}')$ denotes the rest-frame polarization of the initial (final) state, and $u(P, \Lambda)$ is the spinor. The gravitational form factors $A(t)$, $B(t)$ and $C(t)$ can be related to leading-twist
GPDs, which are accessible in exclusive processes~\cite{Ji:2004gf}. Meanwhile, the form factor $\bar{C}(t)$, obtainable from the trace of the energy-momentum tensor, is related to the $\sigma_{\pi N}$ term extracted from pion-nucleon scattering amplitudes~\cite{Alarcon:2011zs,Hoferichter:2015dsa}.

On the other hand, the matrix elements of the quark spin operator 
\begin{align}
S^{\mu\alpha\beta}_{q}(y)=\frac{1}{2}\,\varepsilon^{\mu\alpha\beta\lambda}\,\overline{\psi}(y)\gamma_{\lambda}\gamma_{5}\psi(y)\,,
\end{align}
with $\psi(y)$ and $\bar\psi(y)$ being the quark field are parametrized as 
\begin{equation} 
\langle P', {\bf\Lambda}'\lvert S^{\mu\alpha\beta}_q(0)\rvert P, {\bf\Lambda}\rangle =\frac{1}{2}\,\varepsilon^{\mu\alpha\beta\lambda}\,\overline{u}(P', {\bf\Lambda}')\left[\gamma_{\lambda}\gamma_5\, G^q_{A}(t)+\frac{\Delta_{\lambda}\gamma_5}{2M}\,G^q_{P}(t)\right]u(P, {\bf\Lambda})\,, \label{spar}
\end{equation}
where $G^q_{A}(t)$ and $G^q_{P}(t)$ are the axial vector and pseudoscalar form factors, respectively and the convention $\varepsilon^{0123}=+1$. According to Refs.~\cite{Bakker:2004ib,Leader:2013jra}, the axial form factor is connected to the gravitational form factor associated with the antisymmetric part of the quark EMT, 
$
D_q(t)=-G^q_A(t). 
$
The axial form factor is measurable from quasi-elastic neutrino scattering and pion electroproduction processes~\cite{Bernard:2001rs}. The different angular momentum distributions can thus be defined through the combination of the gravitational form factors and the axial form factor.
\subsection{Distributions in the transverse plane on the light front}
In the light-front (LF) formalism, the impact-parameter distributions of kinetic OAM and spin in the Drell-Yan (DY) frame
are given by~\cite{Lorce:2017wkb} 
\begin{align}
\langle L^{z}\rangle({b}_\perp)&=-i\varepsilon^{3jk}\int\frac{\ud^{2}\vec{\Delta}_\perp}{(2\pi)^2}\,e^{-i\vec{\Delta}_\perp\cdot\vec{b}_\perp}\left.\frac{\partial\langle T^{+k}\rangle}{\partial\Delta^{j}_\perp}\right|_\text{DY}\nonumber\\
&=\Lambda^z\int\frac{\ud^2\vec{\Delta}_\perp}{(2\pi)^2}\,e^{-i\vec{\Delta}_\perp\cdot\vec{b}_\perp}\left[L(t)+t\,\frac{\ud L(t)}{\ud t}\right]_{t=-\vec \Delta^2_\perp},\label{eq:AM}\\
\langle S^{z}\rangle({b}_\perp)&=\frac{1}{2}\,\varepsilon^{3jk}\int\frac{\ud^{2}\vec{\Delta}_\perp}{(2\pi)^2}\,e^{-i\vec{\Delta}_\perp\cdot\vec{b}_\perp}\left.\langle S^{+jk}\rangle\right|_\text{DY} \nonumber\\
&=\frac{\Lambda^z}{2}\int\frac{\ud^{2}\vec{\Delta}_\perp}{(2\pi)^2}\,e^{-i\vec{\Delta}_\perp\cdot\vec{b}_\perp}G_{A}(-\vec \Delta^2_\perp) \; ,\label{eq:spin}
\end{align}
respectively, where 
${2\sqrt{P'^+P^+}}\langle T^{\mu\nu}\rangle\equiv \langle P',S\lvert T^{\mu\nu}(0)\rvert P, S\rangle$  and $L(t)$ is the combination of energy-momentum form factors and the axial form factor, 
\begin{align}
L(t)&=\frac{1}{2}\left[A(t)+B(t)+D(t)\right]=\frac{1}{2}\left[A(t)+B(t)-G_A(t)\right]\,.
\end{align}
 The variable ${\vec b}_\perp$ is the Fourier conjugate to the transverse momentum transfer ${\vec \Delta}_\perp$. The impact parameter ${b}_\perp$ corresponds to the transverse displacement of the active quark from the center of momentum  of the nucleon. 
Meanwhile, the the Belinfante-improved TAM and the total divergence in the impact-parameter are defined as~\cite{Lorce:2017wkb} 
\begin{align}
\langle J^{z}_\text{Bel}\rangle({b}_\perp)&=-i\varepsilon^{3jk}\int\frac{\ud^{2}\vec{\Delta}_\perp}{(2\pi)^2}\,e^{-i \vec{\Delta}_\perp\cdot\vec{b}_\perp}\left.\frac{\partial\langle T^{+k}_\text{Bel}\rangle}{\partial\Delta^{j}_\perp}\right|_\text{DY}\nonumber\\
&=\Lambda^z\int\frac{\ud^2\vec{\Delta}_\perp}{(2\pi)^2}\,e^{-i\vec{\Delta}_\perp\cdot\vec{b}_\perp}\left[J(t)+t\,\frac{\ud J(t)}{\ud t}\right]_{t=-\vec \Delta^2_\perp}, \label{eq:AM_Beli}\\
\langle M^{z}\rangle({b}_\perp)&=\frac{1}{2}\,\varepsilon^{3jk}\int\frac{\ud^{2}\vec{\Delta}_\perp}{(2\pi)^2}\,e^{-i\vec{\Delta}_\perp\cdot\vec{b}_\perp}\,\Delta^l_\perp\left.\frac{\partial\langle S^{l+k}\rangle}{\partial\Delta^j_\perp}\right|_\text{DY}\nonumber\\
&=-\frac{\Lambda^z}{2}\int\frac{\ud^{2}\vec{\Delta}_\perp}{(2\pi)^2}\,e^{-i\vec{\Delta}_\perp\cdot\vec{b}_\perp}\left[t\,\frac{\ud G_{A}(t)}{\ud t}\right]_{t=-\vec \Delta^2_\perp}\,, \label{eq:M2}
\end{align}
respectively, where  
\begin{align}
J(t)&=\frac{1}{2}\left[A(t)+B(t)\right]\,.
\end{align}

Using the two-dimensional Fourier transform of the form factors defined as
\begin{equation}
\tilde{F}(b_\perp)=\int\frac{\ud^2 \vec{\Delta}_\perp}{(2\pi)^2}\,e^{-i\vec\Delta_\perp\cdot\vec b_\perp}\, F(-\vec \Delta^2_\perp) \; , \label{eq:Fourier}
\end{equation}
Eqs.~(\ref{eq:AM})-(\ref{eq:M2}) can be rewritten as 
\begin{align}
\langle L^{z}\rangle({b}_\perp)&=-\frac{\Lambda^z}{2}\,b_{\perp}\,\frac{\ud \tilde{L}(b_{\perp})}{\ud b_{\perp}} \; ,\label{lip}\\
\langle S^{z}\rangle({b}_\perp)&=\frac{\Lambda^z}{2}\,\tilde{G}_{A}(b_{\perp}) \; ,\label{spin}\\
\langle J^{z}_\text{Bel}\rangle({b}_\perp)&=-\frac{\Lambda^z}{2}\,b_{\perp}\,\frac{\ud \tilde{J}(b_{\perp})}{\ud b_{\perp}} \; ,\label{JBel}\\
\langle M^{z}\rangle({b}_\perp)&=\frac{\Lambda^z}{2}\left[\tilde{G}_{A}(b_{\perp})+\frac{1}{2}\,b_{\perp}\frac{\ud\tilde{G}_A(b_{\perp})}{\ud b_{\perp}}\right] \; .\label{divb}
\end{align}
The total angular momentum density $\langle J^{z}\rangle({b}_\perp)$ is then given by
\begin{align}
\langle J^z\rangle(b_\perp)=\langle L^{z}\rangle({b}_\perp)+ \langle S^{z}\rangle({b}_\perp)=\langle J^{z}_\text{Bel}\rangle({b}_\perp)+ \langle M^{z}\rangle({b}_\perp)\,, 
\end{align}
which is different form the ``naive'' density, which is defined by the two-dimensional Fourier transform of $J(t)$,
\begin{align}
 \langle J^z_{\text{naive}}\rangle (b_{\perp})=\Lambda^z\tilde{J}(b_{\perp})\,,
 \label{naive}
\end{align}
 by a correction term
\begin{align}
\langle J^{z}_{\text{corr}}\rangle({b}_\perp)=-\Lambda^z\left[\tilde{L}(b_{\perp})+\frac{1}{2}\,b_{\perp}\,\frac{\ud \tilde{L}(b_{\perp})}{\ud b_{\perp}}\right] \; .
\label{corrb}
\end{align}

Beside the densities mentioned above, the Belinfante-improved TAM can also be formulated as the sum of monopole and quadrupole contributions~\cite{Lorce:2017wkb}
\begin{align}
\langle J_{\text{Bel}}^{z(\text{mono})}\rangle(b_{\perp})&=\frac{\Lambda^z}{3}\left[\tilde{J}(b_{\perp})-b_{\perp}\,\frac{\ud\tilde{J}(b_{\perp})}{\ud b_{\perp}}\right] \;,\label{lbur} \\ 
\langle J_{\text{Bel}}^{z(\text{quad})}\rangle(b_{\perp})&=-\frac{\Lambda^z}{3}\left[\tilde{J}(b_{\perp})+\frac{1}{2}\,b_{\perp}\,\frac{\ud \tilde{J}(b_{\perp})}{\ud b_{\perp}}\right] \; .\label{lquadrup}
\end{align}
The monopole contribution, Eq.~(\ref{lbur}), is the projection of the  expression used by Polyakov and collaborators~\cite{Polyakov:2002yz,Goeke:2007fp} onto the transverse plane. This has later been studied as Polyakov–Goeke distribution in Ref.~\cite{Adhikari:2016dir}. The quadrupole contribution, Eq.~(\ref{lquadrup}), is also the 2D projection of the 3D quadrupole contribution to the Belinfante-improved TAM~\cite{Lorce:2017wkb}, which arises from the breaking of spherical symmetry down
to axial symmetry due to the polarization of the state.

Note that the total divergence (Eq.~\eqref{divb}), the correction (Eq.~\eqref{corrb}), and the quadrupole (Eq.~\eqref{lquadrup})  terms vanish when they are integrated over $\vec b_\perp$. This clarifies how different definitions lead to the same integrated total angular momentum though they are distinct from each other at the density level~\cite{Lorce:2017wkb,Kumar:2017dbf,Adhikari:2016dir}.
\subsection{Generalized parton distributions}
In general, the GPDs are defined through the off-forward matrix elements of the bilocal operators between hadronic states. The unpolarized and helicity dependent quark GPDs for the nucleon are parameterized as~\cite{Ji:1998pc}
\begin{align}
&\int\frac{\ud y^-}{8\pi} e^{ixP^+y^-/2}\braket{P^{\prime},\Lambda^{\prime}|\bar{\psi}(0)\gamma^+\psi(y)|P,\Lambda}|_{y^+=\vec{y}_{\perp}=0}\nonumber\\
&\quad\quad\quad\quad\quad\quad=\frac{1}{2\bar{P}^+} \bar{u}(P^{\prime},\Lambda^{\prime})\left[H^q(x,\zeta,t)\gamma^+
 + E^q(x,\zeta,t) \frac{i\sigma^{+j}\Delta_j}{2M}\right]u(P,\Lambda)\,,\label{HE}\\
&\int\frac{\ud y^-}{8\pi} e^{ixP^+y^-/2}\braket{P^{\prime},\Lambda^{\prime}|\bar{\psi}(0)\gamma^+\gamma_5\Psi(y)|P,\Lambda}|_{y^+=\vec{y}_{\perp}=0}\nonumber\\
&\quad\quad\quad\quad\quad\quad=\frac{1}{2\bar{P}^+} \bar{u}(P^{\prime},\Lambda^{\prime})\Big[\widetilde{H}^q(x,\zeta,t)\gamma^+\gamma_5
 + \widetilde{E}^q(x,\zeta,t) \frac{\gamma_5\Delta^+}{2M}\Big]u(P,\Lambda)\,.\label{HtEt}
\end{align}
Here $H$ and $E$ are the unpolarized quark GPDs, whereas $\widetilde{H}$ and $\widetilde{E}$ represent the helicity dependent GPDs. The kinematical variables are
$\bar{P}=(P'+P)/2$, $\Delta=P'-P$, $\zeta=-\Delta^+/2\bar{P}^+$
and $t=\Delta^2$. For $\zeta=0$, $t=-\vec{\Delta}_\perp^2$. We consider the light cone gauge $A^+ = 0$, which indicates that the gauge-link between the quark
fields in Eqs.~(\ref{HE}) and (\ref{HtEt}) is unity. In this paper, we concentrate only on the GPDs relevant to the angular momentum densities, i.e., $H$, $E$ and $\widetilde{H}$ at the zero skewness limit. Note that one has to consider nonzero skewness to compute GPD $\widetilde{E}$, which is not needed for this work.

Substituting the nucleon states within the valence Fock sector 
\begin{align}
\ket{P,{\Lambda}} =&  \int \prod_{i=1}^{3} \left[\frac{{\rm d}x_i{\rm d}^2 \vec{k}_{i\perp}}{\sqrt{x_i}16\pi^3}\right] 16\pi^3\delta \left(1-\sum_{i=1}^{3} x_i\right) \delta^2 \left(\sum_{i=1}^{3}\vec{k}_{i\perp}\right)\nonumber\\
&\times \Psi^{\Lambda}_{\{x_i,\vec{k}_{i\perp},\lambda_i\}} \ket{\{x_iP^+,\vec{k}_{i\perp}+x_i\vec{P}_{\perp},\lambda_i\}}\,,\label{wavefunction_expansion}
\end{align}
and the quark field operators in  Eqs.~(\ref{HE}) and (\ref{HtEt}) leads to the GPDs in terms of the overlap of the LFWFs
\begin{align}
H^q(x,0,t)=& 
 \sum_{\{\lambda_i\}} \int \left[{\rm d}\mathcal{X} \,{\rm d}\mathcal{P}_\perp\right]\, \Psi^{\uparrow *}_{\{x^{\prime}_i,\vec{k}^{\prime}_{i\perp},\lambda_i\}}\Psi^{\uparrow}_{\{x_i,\vec{k}_{i\perp},\lambda_i\}} \delta(x-x_1)\,,  \label{eq:H}  \\
E^q(x,0,t)=& -\frac{2M}{(q^1-iq^2)}
 \sum_{\{\lambda_i\}} \int \left[{\rm d}\mathcal{X} \,{\rm d}\mathcal{P}_\perp\right]\, \Psi^{\uparrow *}_{\{x^{\prime}_i,\vec{k}^{\prime}_{i\perp},\lambda_i\}}\Psi^{\downarrow}_{\{x_i,\vec{k}_{i\perp},\lambda_i\}} \delta(x-x_1)\,, \label{eq:E}\\
 \widetilde{H}^q(x,0,t)=& 
 \sum_{\{\lambda_i\}} \int \left[{\rm d}\mathcal{X} \,{\rm d}\mathcal{P}_\perp\right]\, \lambda_1\, \Psi^{\uparrow *}_{\{x^{\prime}_i,\vec{k}^{\prime}_{i\perp},\lambda_i\}}\Psi^{\uparrow}_{\{x_i,\vec{k}_{i\perp},\lambda_i\}} \delta(x-x_1)\,,   \label{eq:Ht} 
\end{align}
where 
\begin{align}
\left[{\rm d}\mathcal{X} \,{\rm d}\mathcal{P}_\perp\right]=\prod_{i=1}^3 \left[\frac{{\rm d}x_i{\rm d}^2 \vec{k}_{i\perp}}{16\pi^3}\right] 16 \pi^3 \delta \left(1-\sum_{i=1}^{3} x_i\right) \delta^2 \left(\sum_{i=1}^{3}\vec{k}_{i\perp}\right),  
\end{align}
and the light-front momenta are $x^{\prime}_1=x_1$; $\vec{k}^{\prime}_{1\perp}=\vec{k}_{1\perp}+(1-x_1)\vec{\Delta}_{\perp}$ for the struck quark ($i=1$) and $x^{\prime}_i={x_i}; ~\vec{k}^{\prime}_{i\perp}=\vec{k}_{i\perp}-{x_i} \vec{\Delta}_{\perp}$ for the spectators ($i\ne1$) and $\lambda_1=1~(-1)$ for the struck quark helicity. The proton helicity  is designated by $\Lambda=\uparrow(\downarrow)$, where $\uparrow$ and $\downarrow$ correspond to $+1$ and $-1$, respectively.

Integrating the non-local matrix element that parameterized the GPDs, over $x$ leads to the local matrix elements yielding the form factors. In the Drell-Yan frame, the expressions for the form factors are very similar to the expressions for GPDs, except that the longitudinal momentum fraction $x$ of the struck quark is not integrated out in the  GPDs' expressions. Thus, GPDs defined in Eqs.~(\ref{eq:H}), (\ref{eq:E}) and (\ref{eq:Ht}) are also known as momentum-dissected form factors and measure the contribution of the struck quark with momentum fraction $x$ to the corresponding form factors. The electromagnetic form factors are related to the first moments of the unpolarized GPDs for the nucleon by the sum rules on the light-front as
\begin{align}
 F_1^q(t)=&\int \ud x \, H^q(x,0,t)\,;\quad
 F_2^q(t)=\int \ud x \, E^q(x,0,t)\,,\label{eq:FF}
\end{align}
where $F_1^q(t)$ and $F_2^q(t)$ are the Dirac (charge) and the Pauli (magnetic) form factors, respectively whereas, the axial form factor is connected to the helicity dependent GPD as
\begin{align}
G_A^q(t)=&\int dx \, \widetilde{H}^q(x,0,t)\,.\label{eq:gA}
\end{align}
Meanwhile, the gravitational form factors which are parameterized through
the matrix elements of the EMT are linked to the second moment of GPDs
\begin{align}
 A^q(t)=\int \ud x \, x\,H^q(x,0,t)\,;\quad\quad
 B^q(t)=\int \ud x \, x\, E^q(x,0,t)\,.\label{eq:GF}
\end{align}

The transverse impact parameter dependent GPDs are obtained via the FT of the GPDs with respect to the momentum transfer along the transverse direction $\vec{\Delta}_\perp$~\cite{Burkardt:2002hr}:
\begin{align}
 {F}(x, {b}_\perp)& =
\int \frac{\ud^2{\vec \Delta}_\perp}{(2\pi)^2}
e^{-i {\vec \Delta}_\perp \cdot {\vec b}_\perp }
F(x,0,t)\,,\label{eq:Hb}
\end{align}
with $F$ being the GPDs $H$, $E$ and $\widetilde{H}$. The $H(x,b_\perp)$ 
provides the description of the density of unpolarized quarks in the unpolarized proton, while $E(x,b_\perp)$ is responsible for a deformation of the density in the transversely polarized proton. 
The transverse distortion can be linked to Ji’s angular momentum relation. Ji has shown that the TAM of quarks and gluons can be expressed in terms of GPDs~\cite{Ji:PRL}:
\begin{align}
J^z=\frac{1}{2} \int \ud x \, x\,[H(x,0,0)+E(x,0,0)]\,.
\end{align}
This sum rule is appropriate at the forward limit of the GPDs and relates the $z$-component of the TAM of the constituents in a nucleon polarized in the $z$-direction only. Again in the impact parameter space, the sum rule has a simple interpretation
for a transversely polarized nucleon~\cite{Burkardt:2005hp}. The term involving $E(x,0,0)$ arises due to the transverse deformation of the distribution in the center of momentum frame, whereas the term containing $H(x,0,0)$ is an overall transverse shift when going from the transversely polarized nucleon in instant form to the front form. Meanwhile, the helicity-dependent GPD $\widetilde{H}$ in the impact parameter space reflects the difference in the density of parton with helicity equal or opposite to the nucleon helicity~\cite{Diehl:2005jf,Boffi:2007yc,Pasquini:2007xz,Mondal:2017wbf}. This GPD has a direct connection with the partonic spin contribution to the TAM of the nucleon.

We can now rewrite the distributions defined in Eqs.~(\ref{lip})-(\ref{lquadrup}) using the impact parameter dependent GPDs, where $\tilde{L}(b_\perp)$, $\tilde{J}(b_\perp)$ and $\tilde{G}_A(b_\perp)$ are given by 
\begin{align}
  \tilde{L}({b}_\perp) &=\frac{1}{2}
\int \ud x \, \left\{x\left[{H}(x,b_\perp)+{E}(x,b_\perp)\right]-\widetilde{{H}}(x,b_\perp)\right\}\,,\label{eq:Lb}\\
  \tilde{J}({b}_\perp) &=\frac{1}{2}
\int \ud x \, x\left[{H}(x,b_\perp)+{E}(x,b_\perp)\right]\,,\label{eq:Jb}\\
  \tilde{G}_A({b}_\perp) &=
\int \ud x \,\widetilde{{H}}(x,b_\perp)\,.\label{eq:Gb}
\end{align}
\begin{figure}
\begin{tabular}{cc}
\subfloat[]{\includegraphics[scale=0.35]{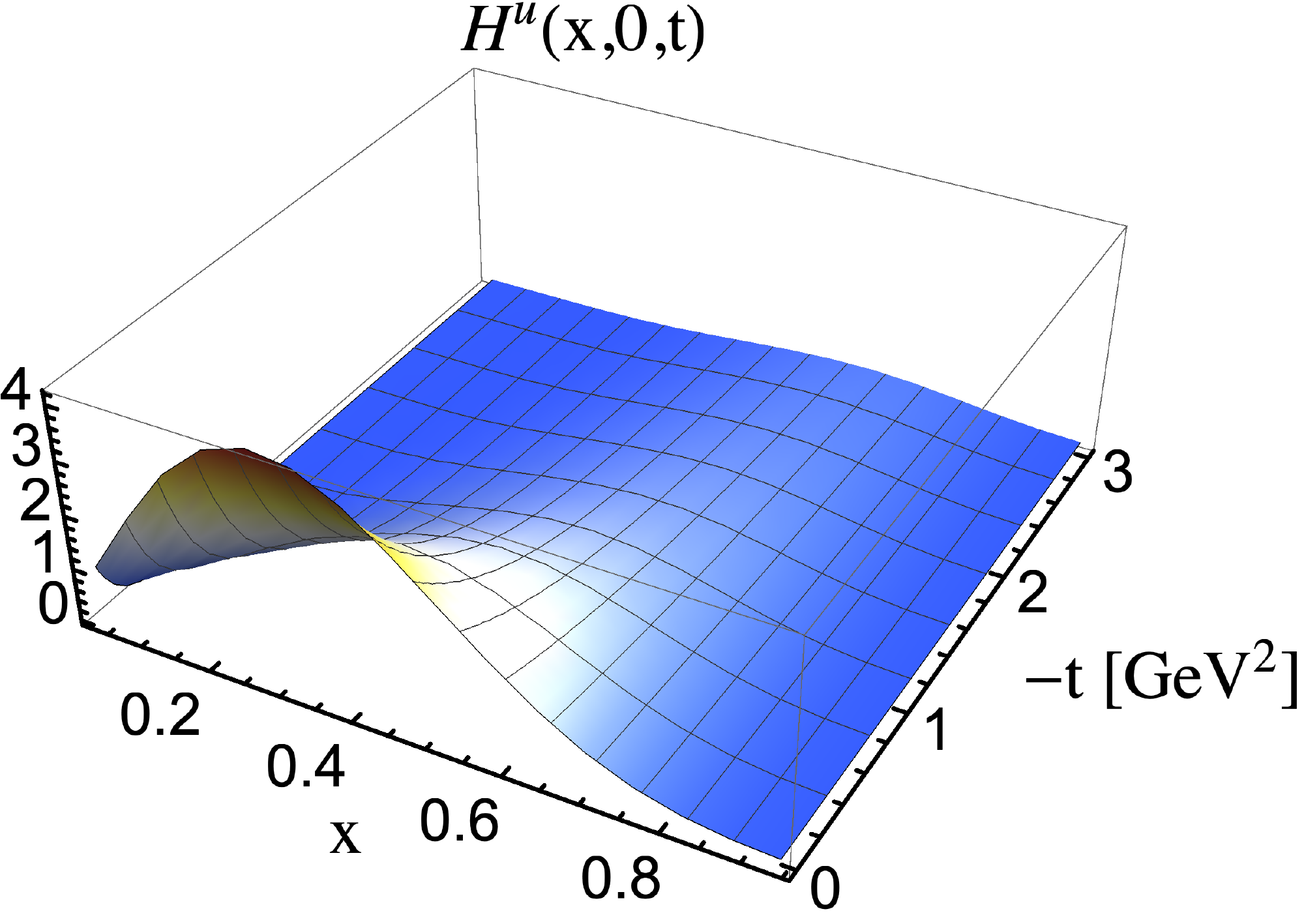}}
\end{tabular}
\begin{tabular}{cc}
\subfloat[]{\includegraphics[scale=0.35]{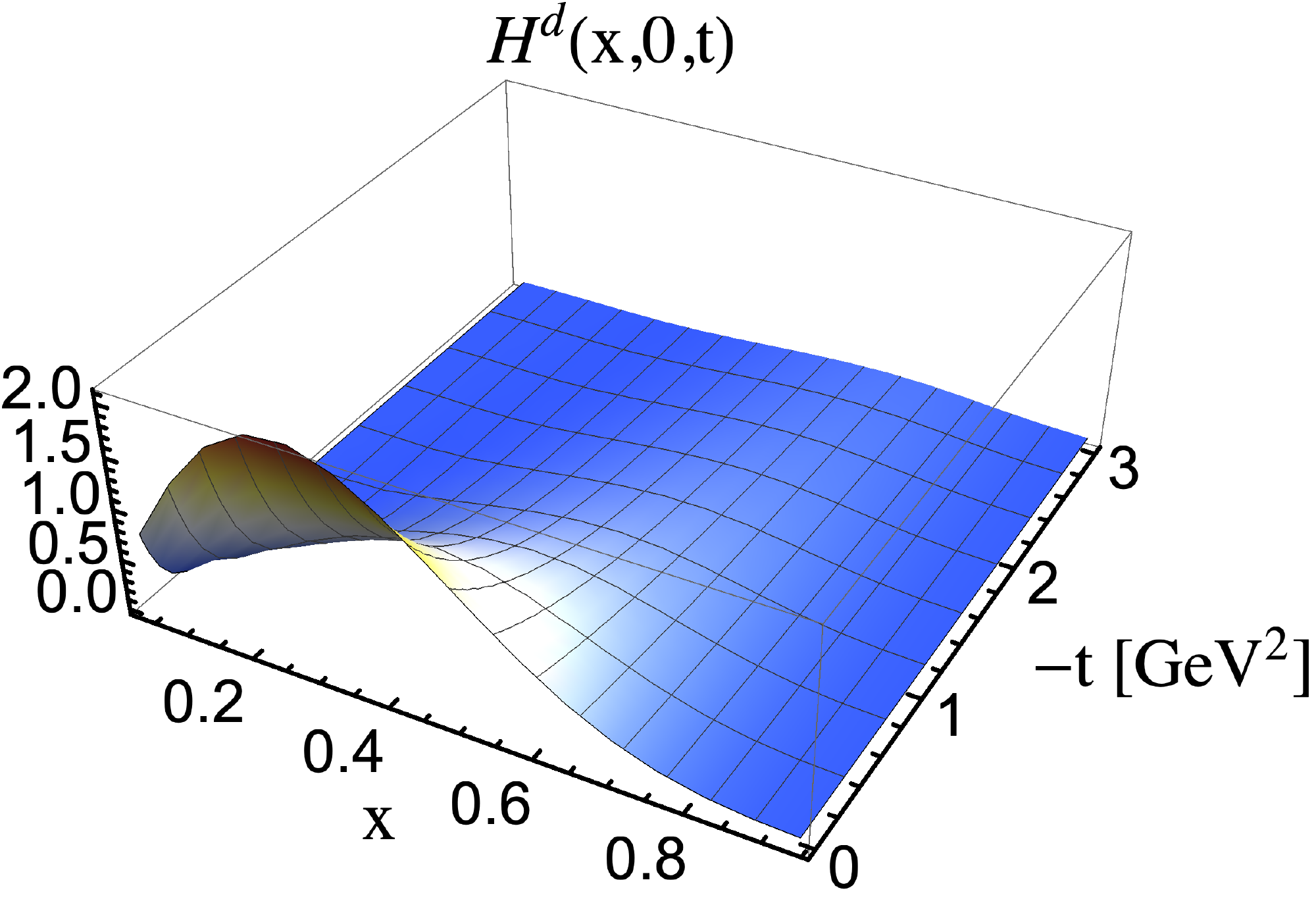}}
\end{tabular}
\begin{tabular}{cc}
\subfloat[]{\includegraphics[scale=0.35]{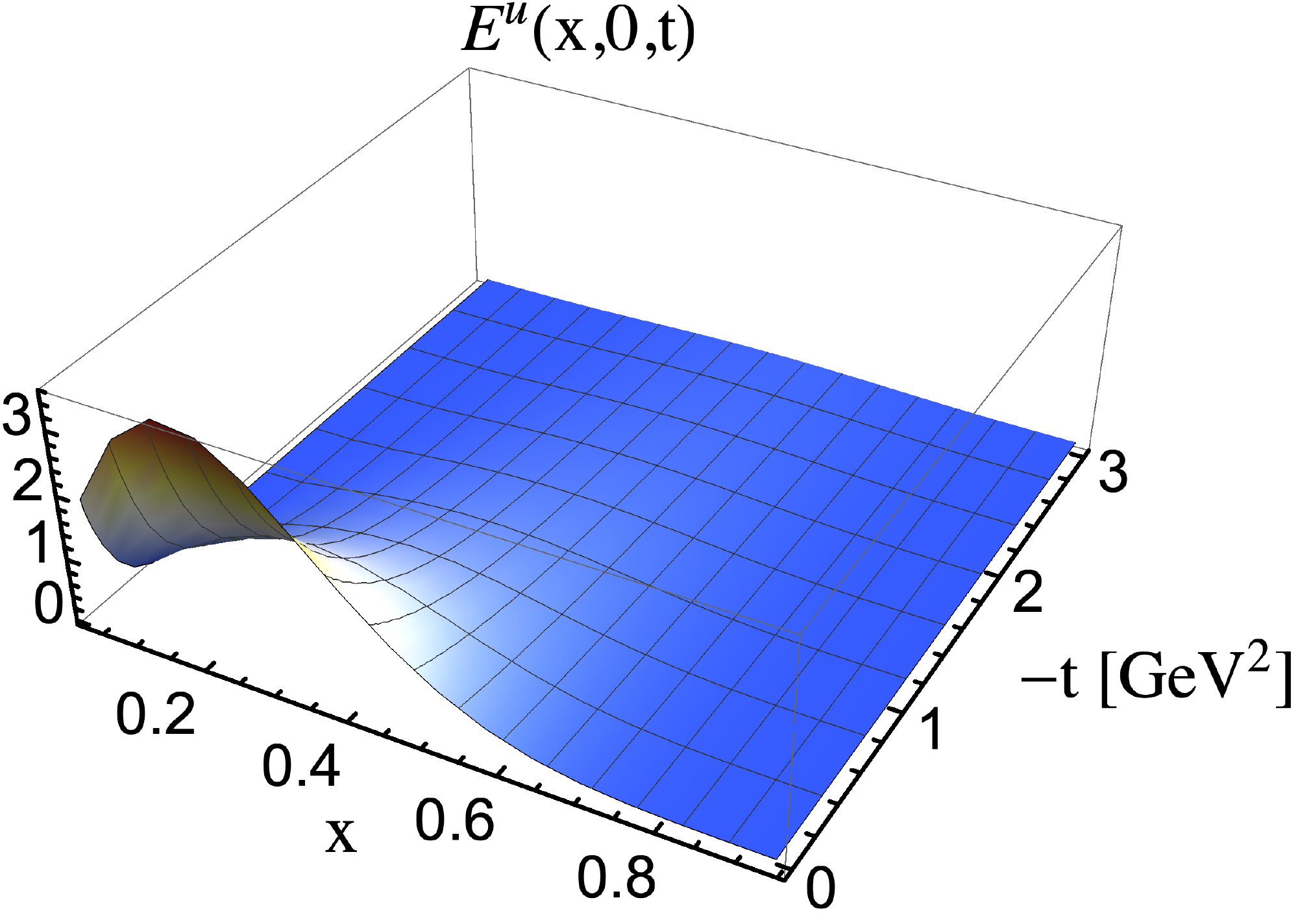}}
\end{tabular}
\begin{tabular}{cc}
\subfloat[]{\includegraphics[scale=0.35]{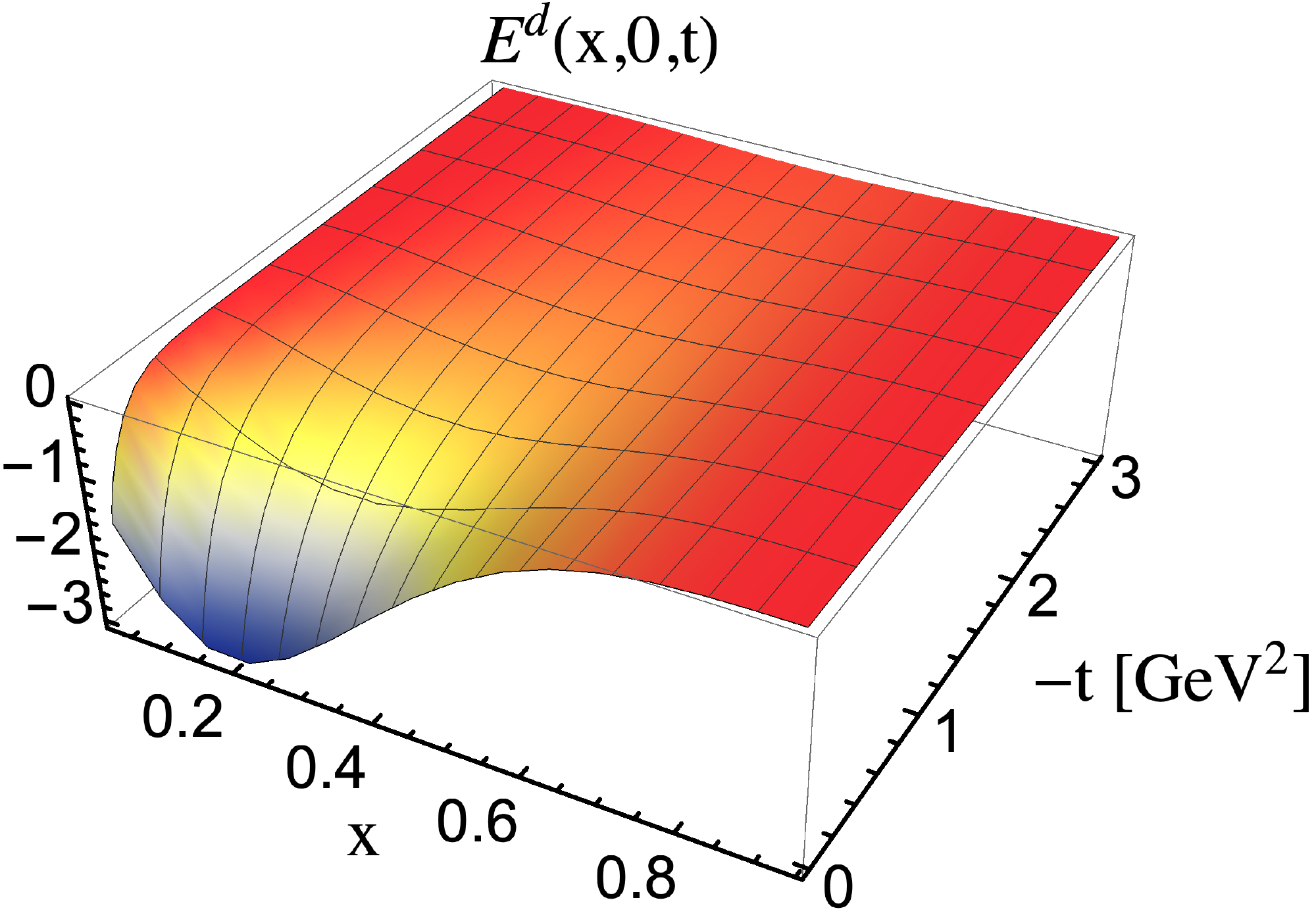}}
\end{tabular}
\begin{tabular}{cc}
\subfloat[]{\includegraphics[scale=0.35]{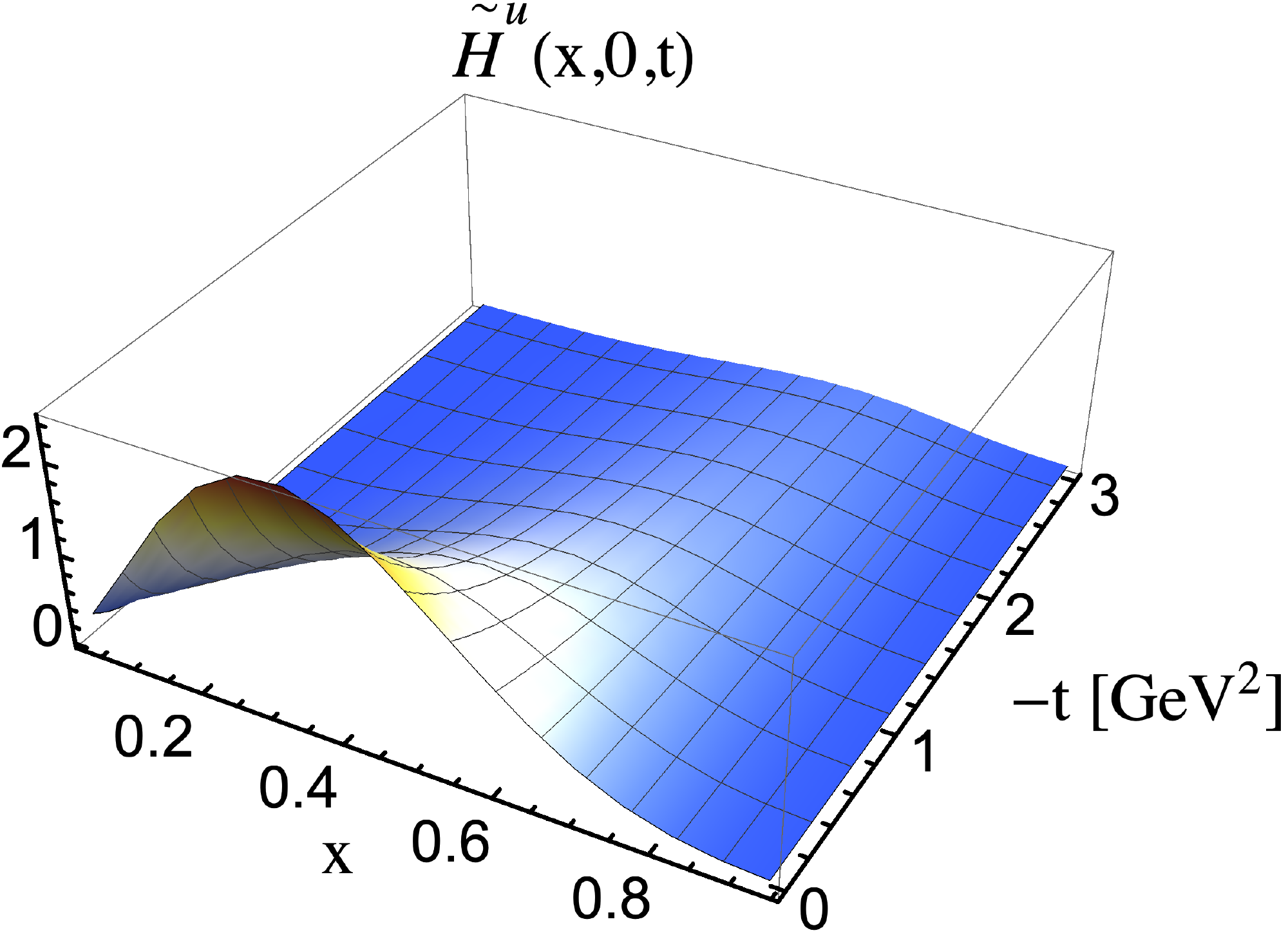}}
\end{tabular}
\begin{tabular}{cc}
\subfloat[]{\includegraphics[scale=0.35]{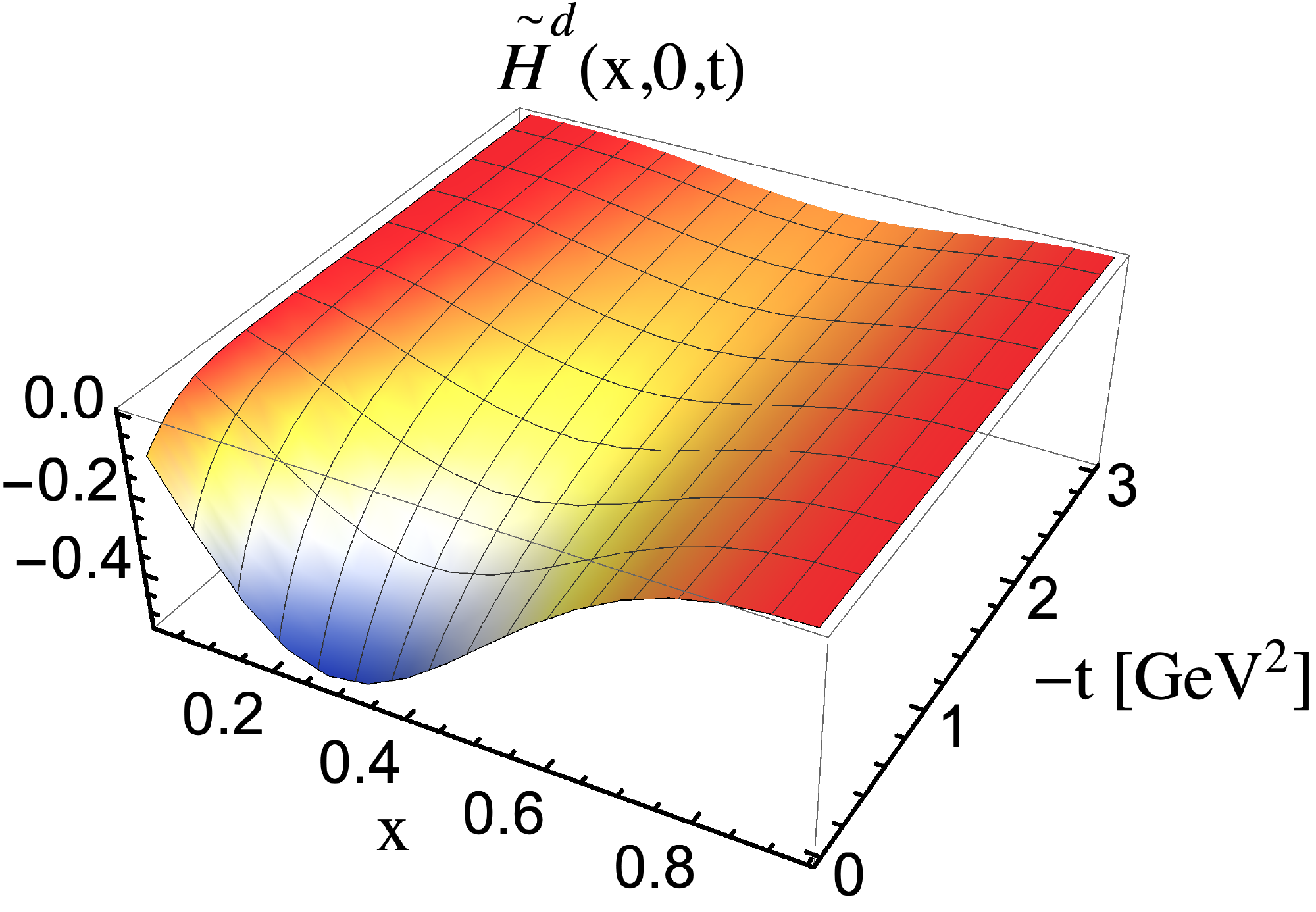}}
\end{tabular}
\caption{The valence quark GPDs of the proton: (a) $H(x,0,t)$, (c) $E(x,0,t)$, and (e) $\widetilde{H}(x,0,t)$ are for the valence up quark; (b), (d) and (f) are same as (a), (c), and (e), respectively, but for the valence down quark as functions of $x$ and $-t$.}
\label{Fig:gpds}
\end{figure}
\begin{figure}
\begin{tabular}{cc}
\subfloat[]{\includegraphics[scale=0.35]{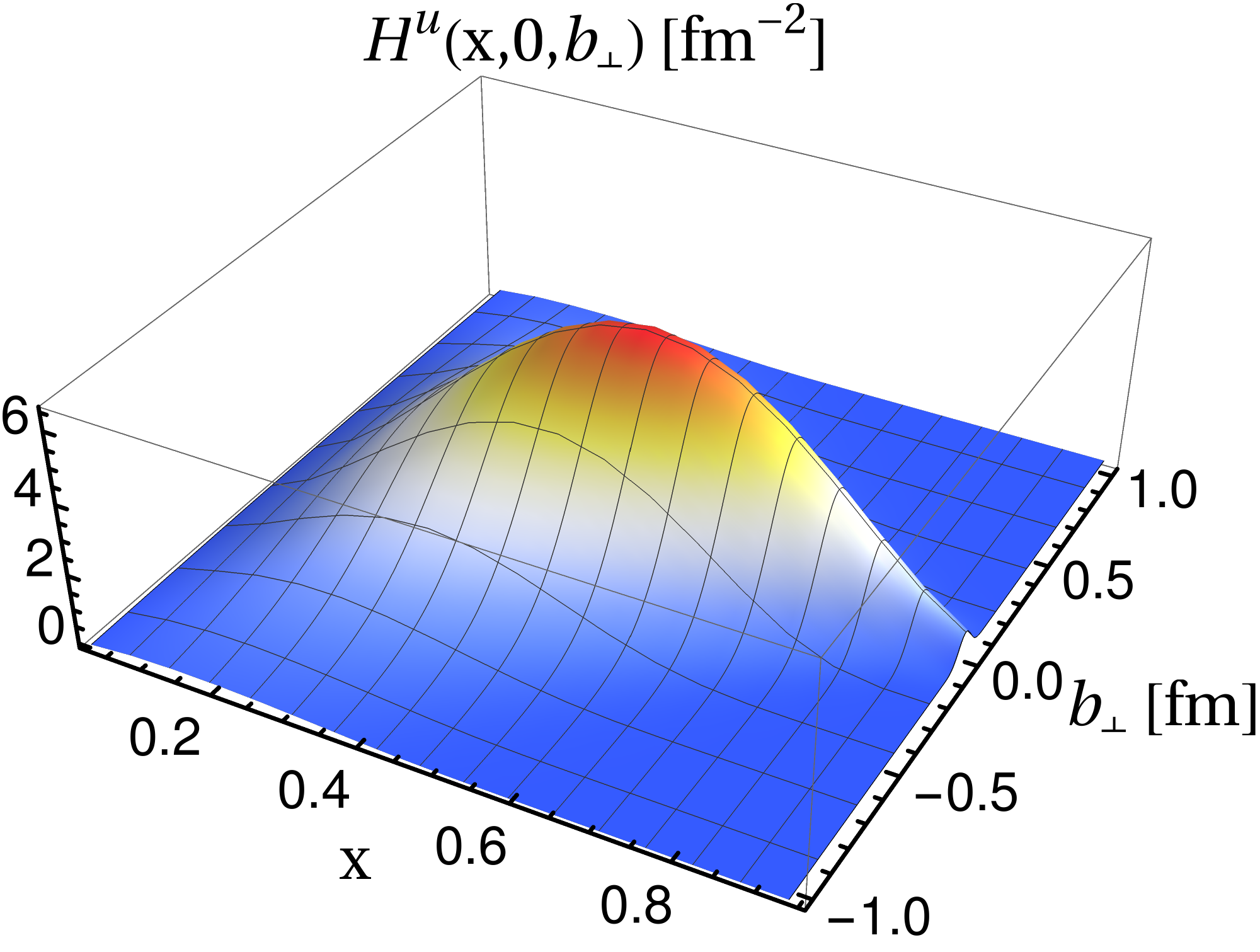}}
\end{tabular}
\begin{tabular}{cc}
\subfloat[]{\includegraphics[scale=0.35]{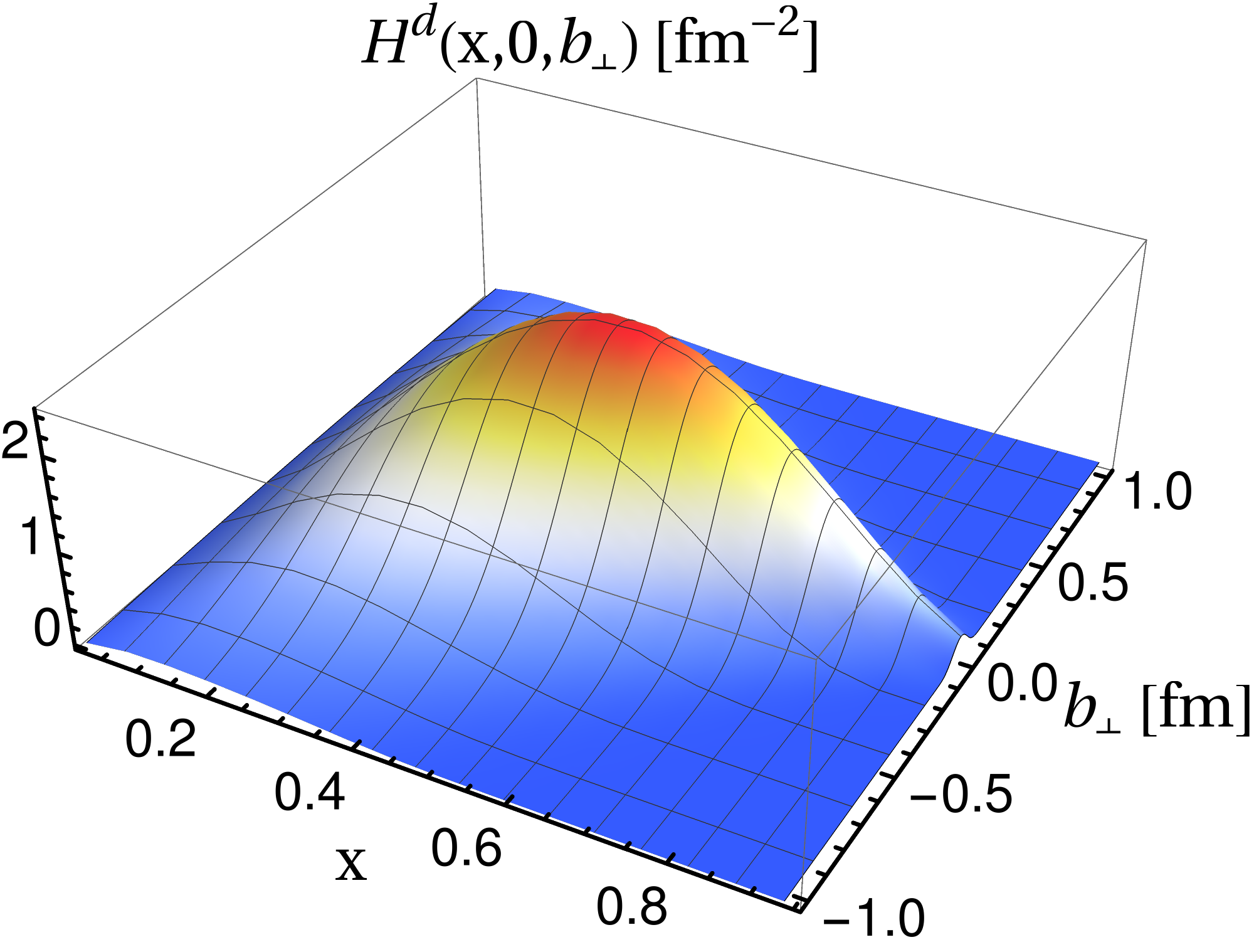}}
\end{tabular}
\begin{tabular}{cc}
\subfloat[]{\includegraphics[scale=0.35]{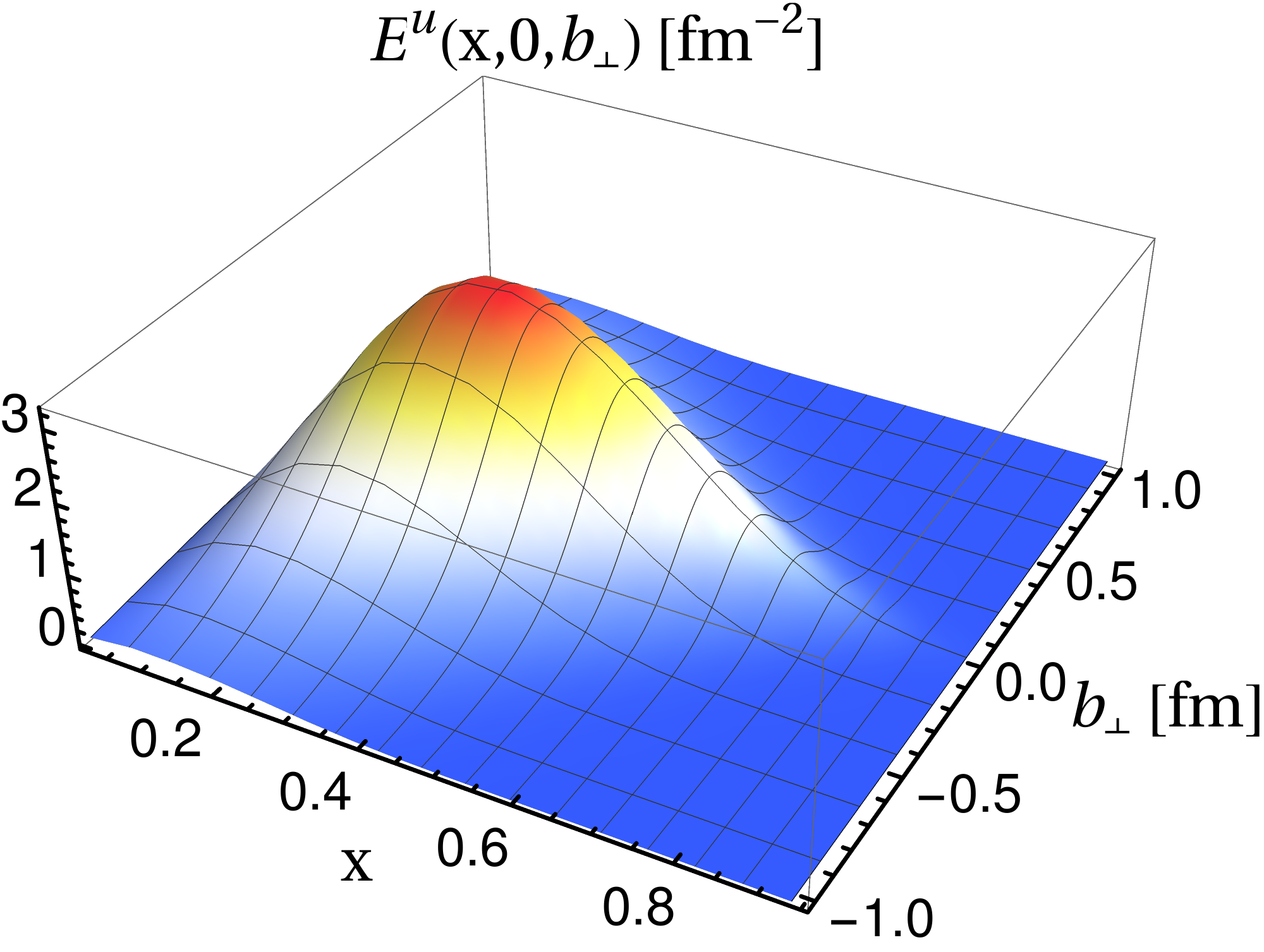}}
\end{tabular}
\begin{tabular}{cc}
\subfloat[]{\includegraphics[scale=0.35]{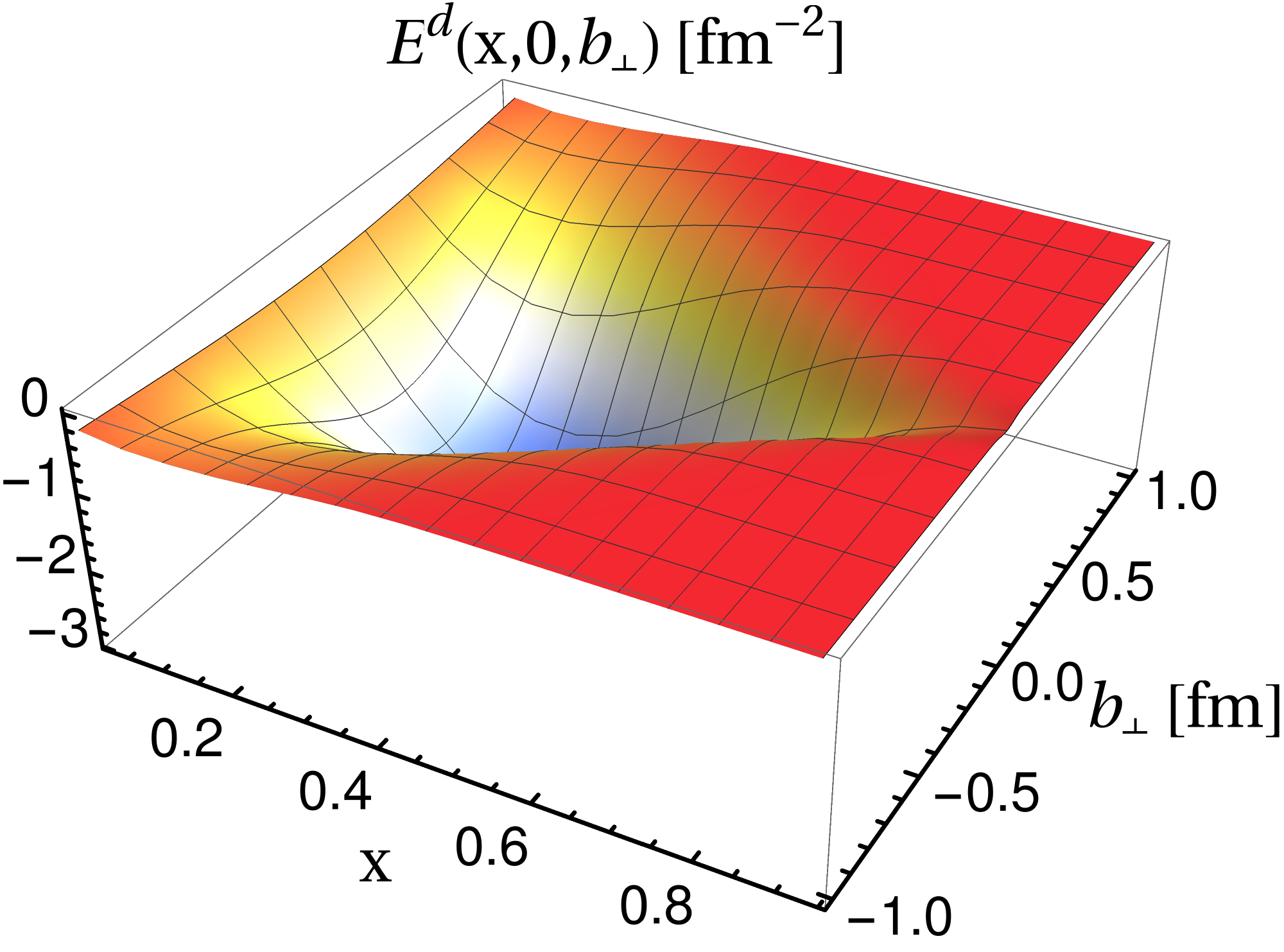}}
\end{tabular}
\begin{tabular}{cc}
\subfloat[]{\includegraphics[scale=0.35]{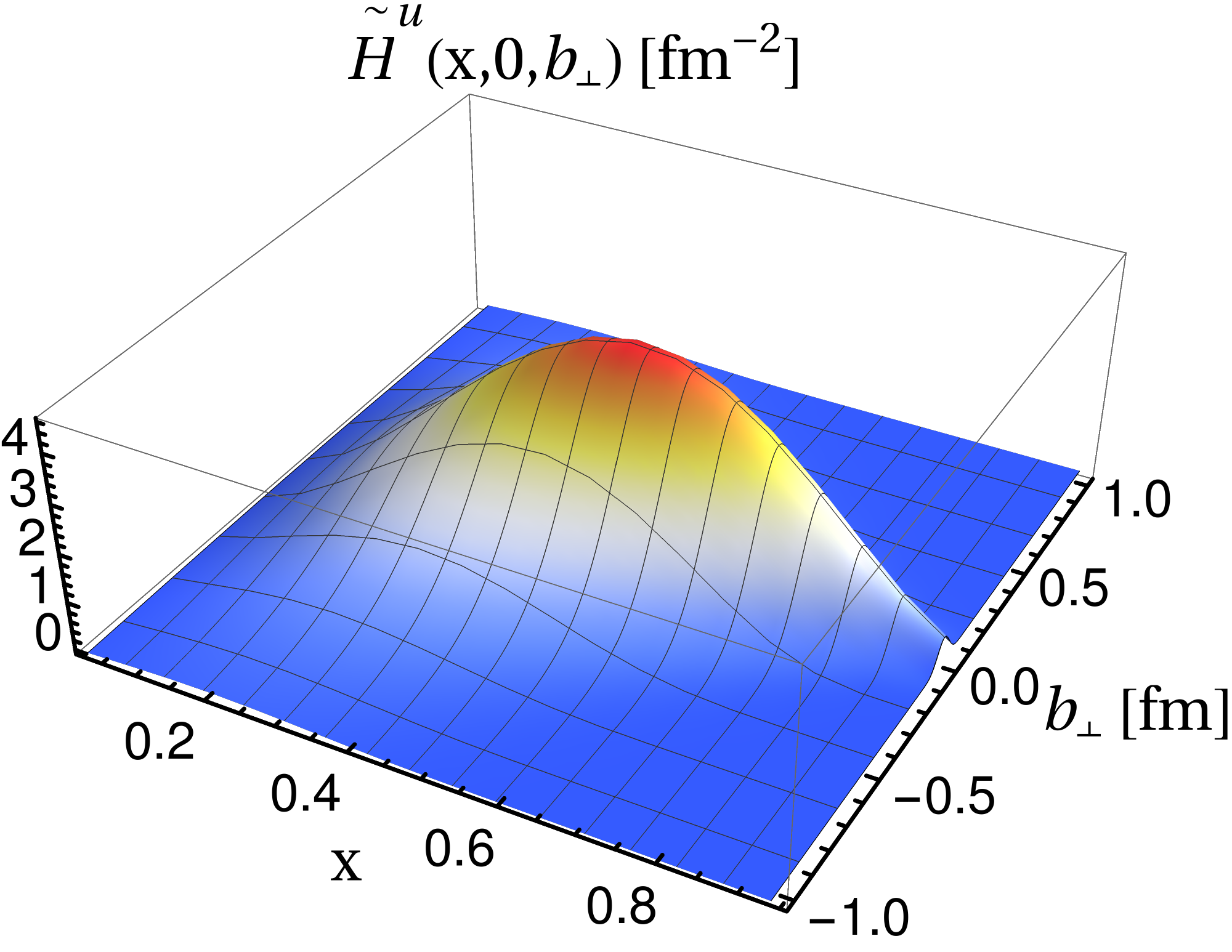}}
\end{tabular}
\begin{tabular}{cc}
\subfloat[]{\includegraphics[scale=0.35]{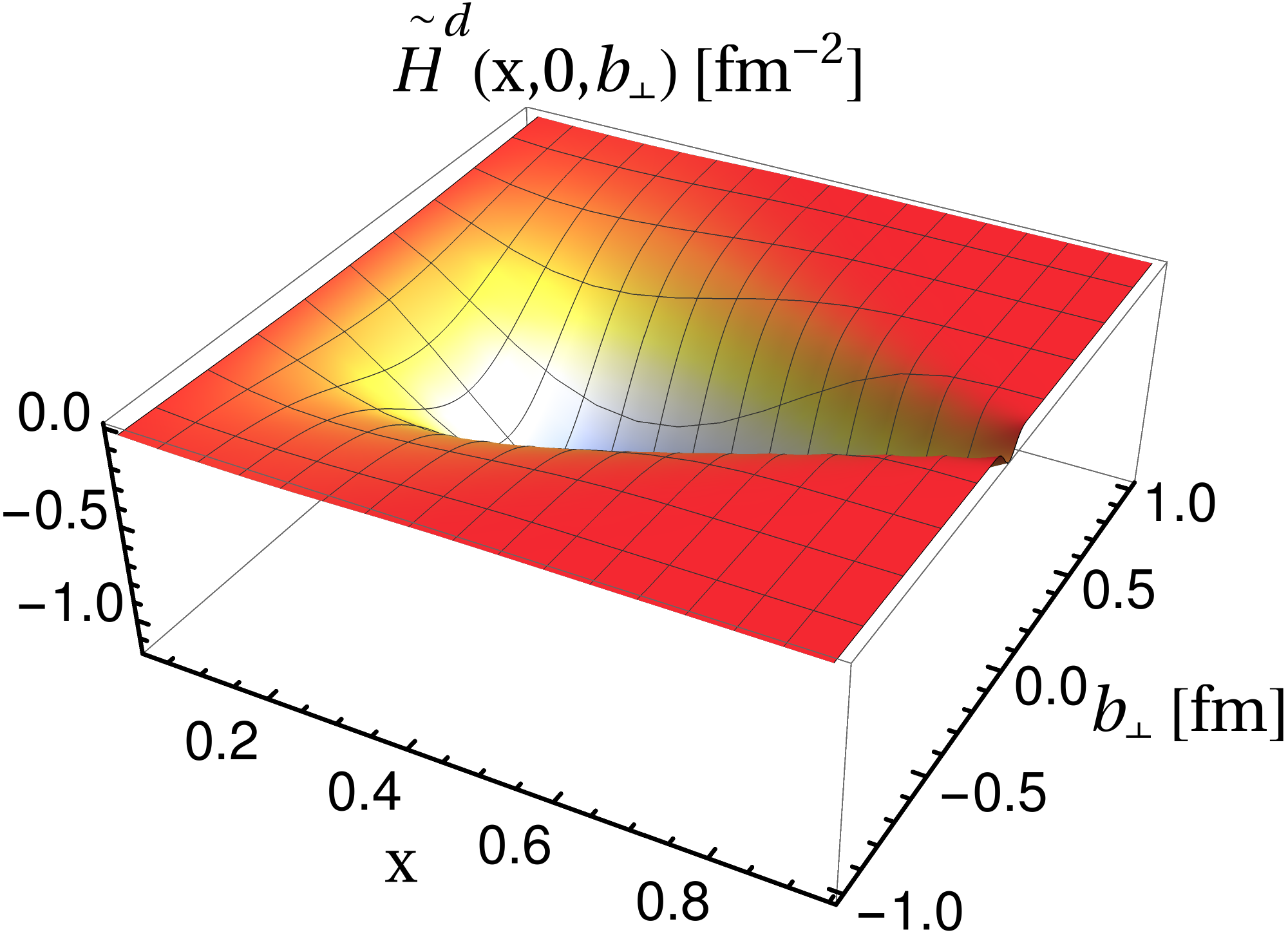}}
\end{tabular}
\caption{The valence quark GPDs of the proton in the transverse impact parameter space: (a) $H(x, b_\perp)$, (c) $E(x, b_\perp)$, and (e) $\widetilde{H}(x, b_\perp)$ are for the valence up quark; (b), (d) and (f) are same as (a), (c), and (e), respectively, but for the valence down quark as functions of $x$ and $b_\perp$.}
\label{Fig:impact_gpds}
\end{figure}
\section{Numerical results and discussions} \label{sec:results}
The LFWFs of the valence quarks in the proton have been solved in the BLFQ framework with the basis truncation $N_{\rm max}=10$ and $K=16.5$ and the model parameters $\{m_{\rm q/KE},~m_{\rm q/OGE},~\kappa,~\alpha_s\}=\{0.3~{\rm GeV},~0.2~{\rm GeV},~0.34~{\rm GeV},~1.1\pm 0.1\}$ and the HO scale parameter $b=0.6$ GeV. 
The parameters in our model are fixed to fit the nucleon mass and the flavor
Dirac form factors~\cite{Mondal:2019jdg}. We estimate an uncertainty on the coupling that accounts for the model selections and major fitting uncertainties. The uncertainty for the $\alpha_s$ decreases with increasing basis cutoffs $N_{\rm{max}}$~\cite{Xu:2021wwj}.  We employ the resulting wave functions to investigate the GPDs for the proton. We insert the valence wave functions given by Eq.~(\ref{wavefunctions}) into Eqs.~(\ref{eq:H}), (\ref{eq:E}) and (\ref{eq:Ht}) to compute the valence quark GPDs inside the proton.

We show the unpolarized GPDs, $H^q$ and $E^q$, and the helicity dependent GPD $\widetilde{H}^q$ as functions of $x$ and $-t$ for the proton in Fig.~\ref{Fig:gpds}. The GPD $E$ in the proton has its peak located at a lower $x$ than the peaks in $H$ and $\widetilde{H}$.  In addition, the GPD E  falls faster than the other two GPDs at large-$x$. Meanwhile, the GPD $\widetilde{H}$ exhibits the similar behavior as manifested by the GPD $H$. This is due to the fact that $E$ involves the overlaps of the wave functions with different orbital angular momentum $L_z=0$ and $L_z=\pm 1$ and the other two GPDs entail the overlaps of the wave functions of the same orbital angular momentum. The magnitudes of distributions decrease and the peaks along $x$ shift towards larger values of $x$ with increasing momentum transfer $-t$ similar to that observed in the other phenomenological models for the nucleon~\cite{Ji:1997gm,Scopetta:2002xq,Petrov:1998kf,Penttinen:1999th,Boffi:2002yy,Boffi:2003yj,Vega:2010ns,Chakrabarti:2013gra,Mondal:2015uha,Chakrabarti:2015ama,Mondal:2017wbf,deTeramond:2018ecg,Xu:2021wwj} as well as for the light mesons~\cite{Kaur:2018ewq,deTeramond:2018ecg,Zhang:2021mtn,Kaur:2020vkq}.

\begin{figure}
\begin{tabular}{cc}
\subfloat[]{\includegraphics[scale=0.38]{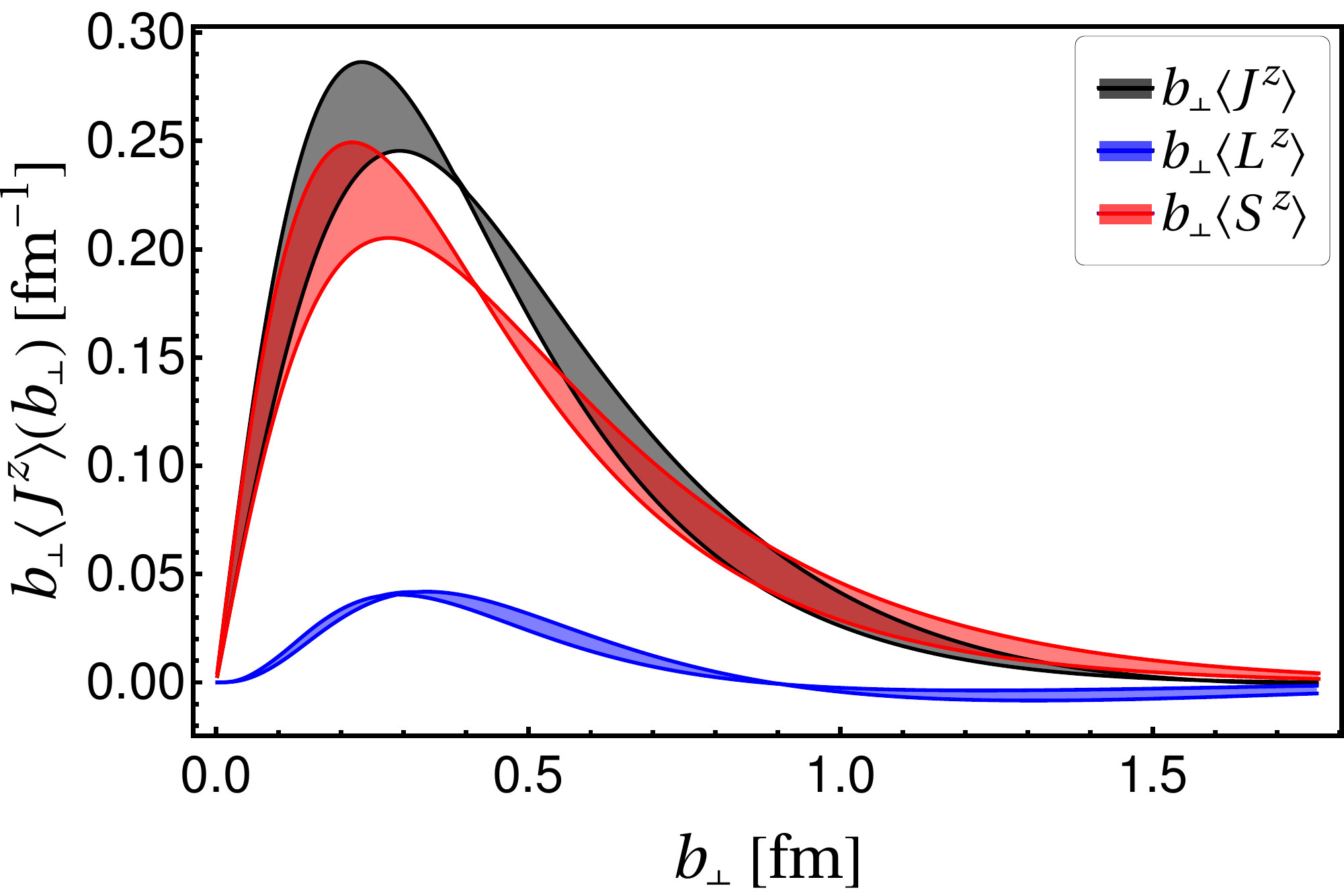}}
\end{tabular}
\begin{tabular}{cc}
\subfloat[]{\includegraphics[scale=0.38]{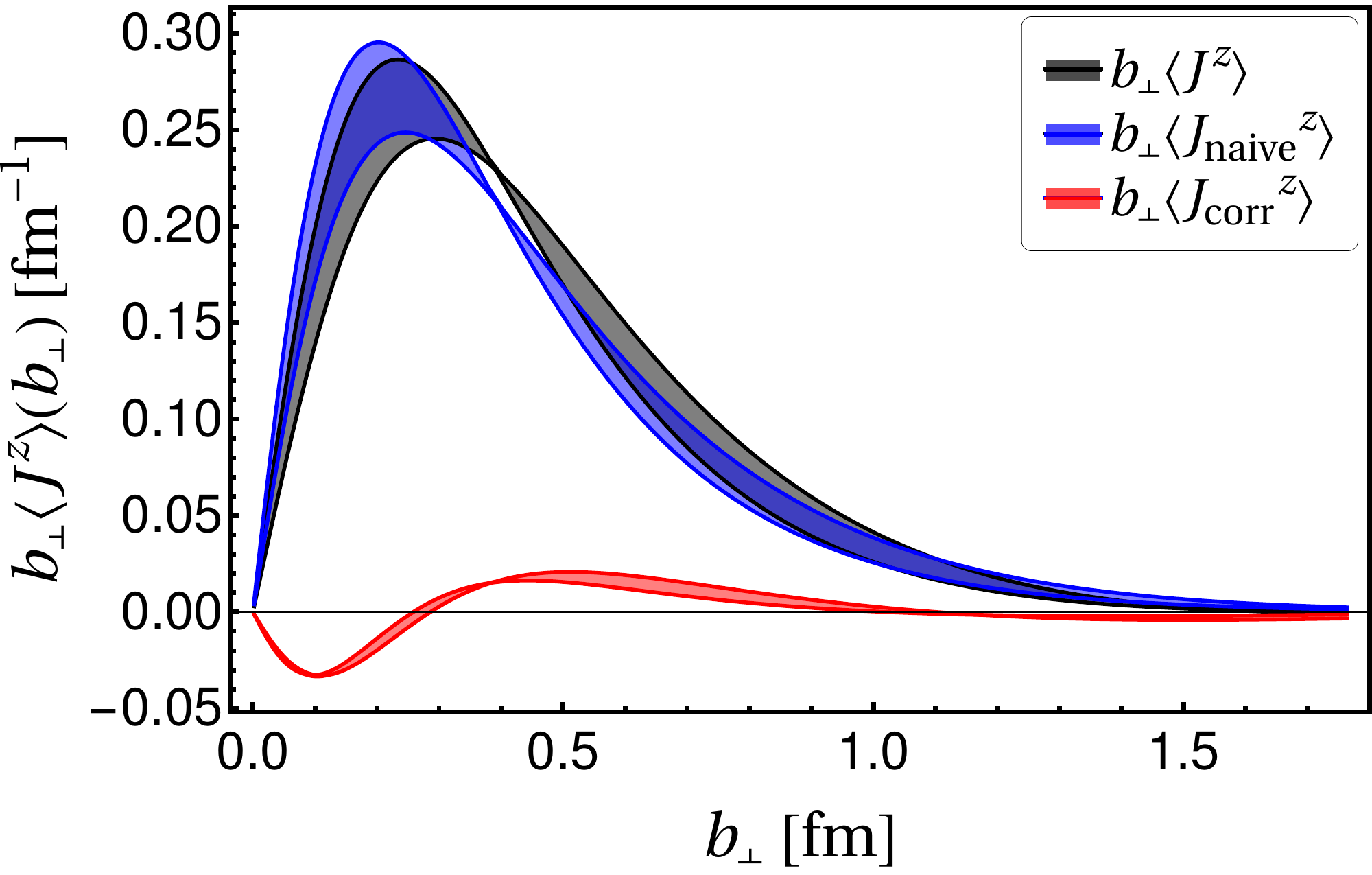}}
\end{tabular}
\begin{tabular}{cc}
\subfloat[]{\includegraphics[scale=0.38]{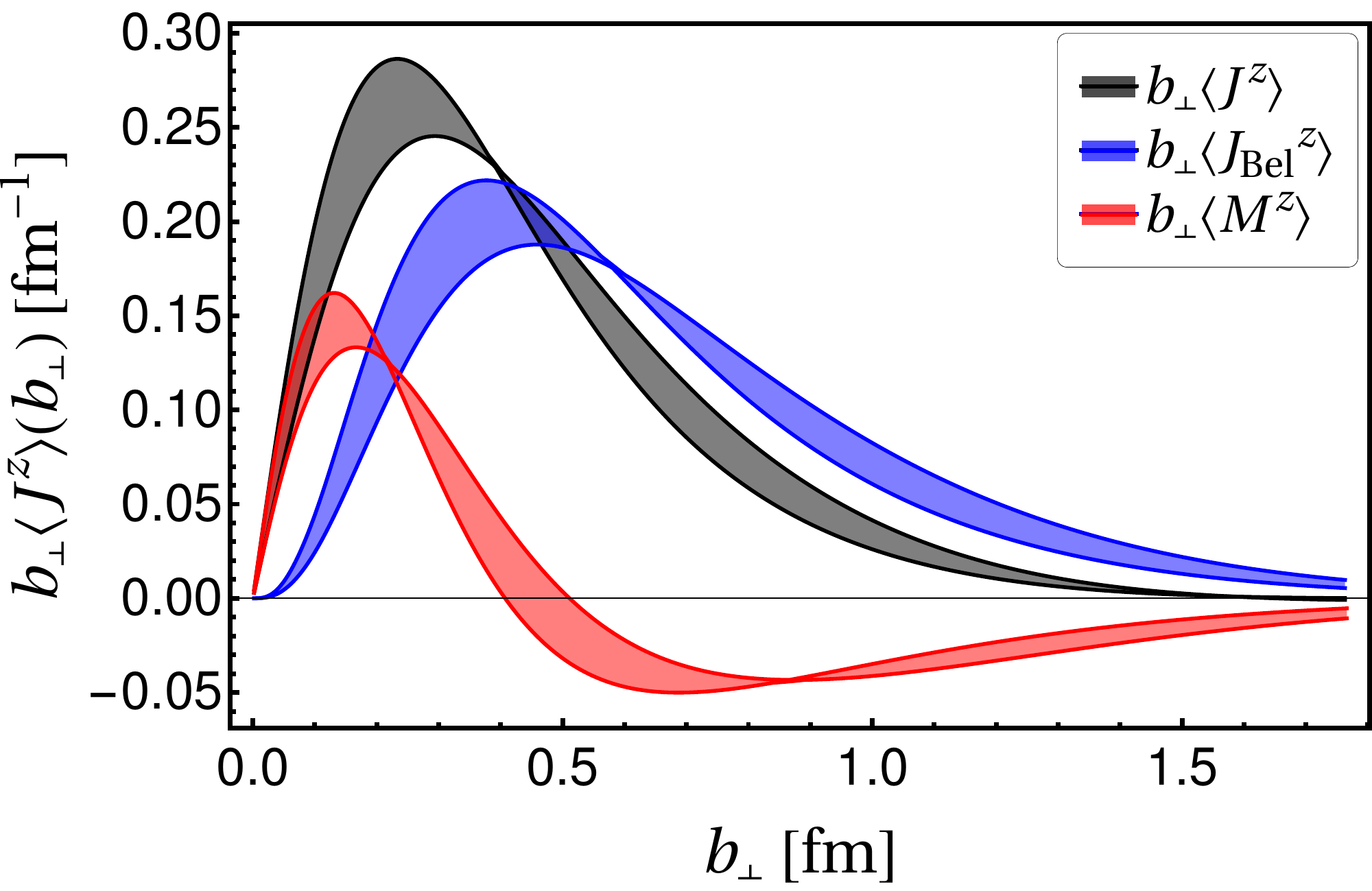}}
\end{tabular}
\begin{tabular}{cc}
\subfloat[]{\includegraphics[scale=0.38]{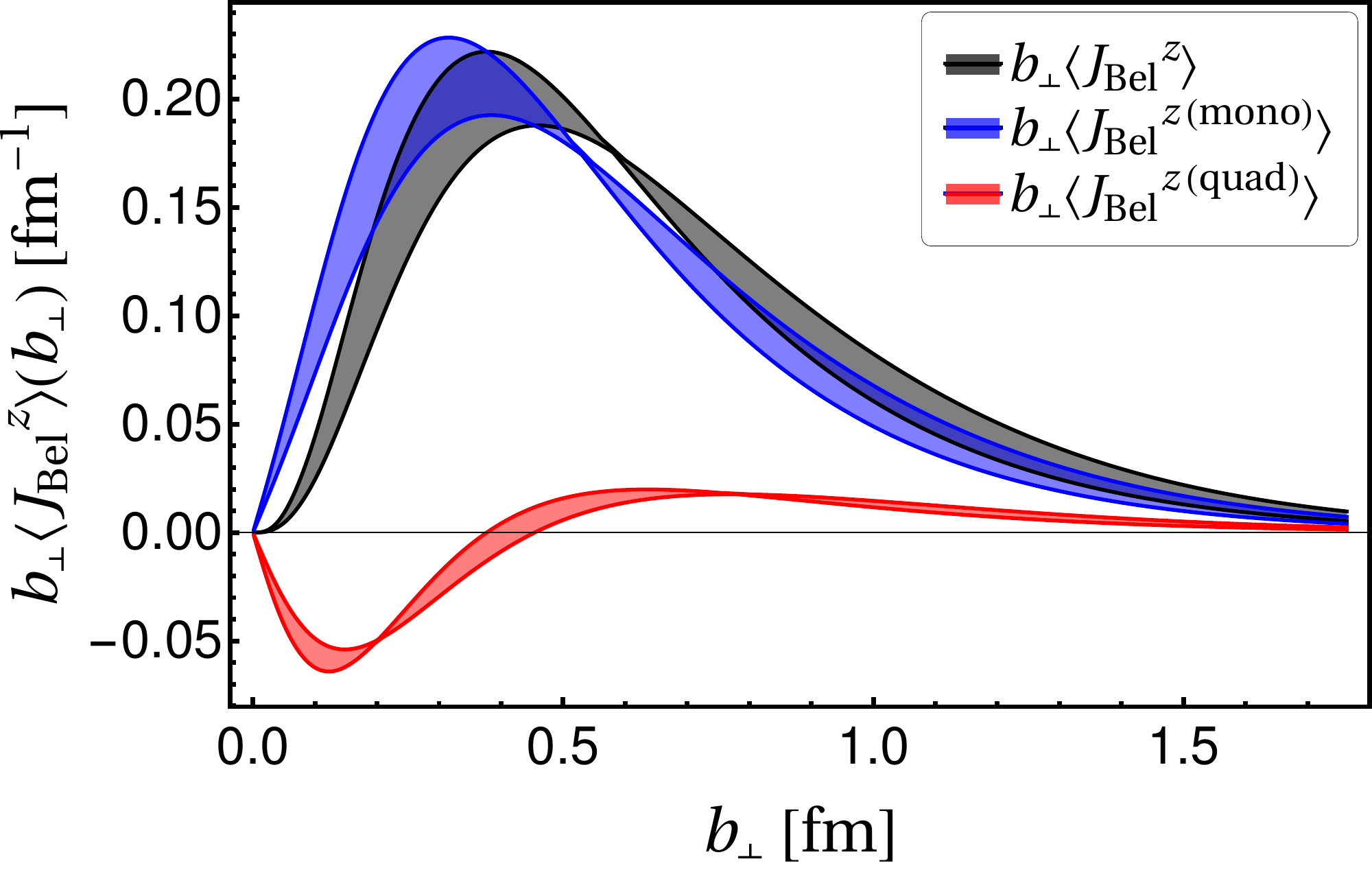}}
\end{tabular}
\caption{Angular momentum distributions summing over the up and the down quark contributions multiplied by $b_\perp$ as functions of $b_\perp$: 
(a) the kinetic TAM density $\langle J^z\rangle(b_\perp)$ (black band) as the sum of the spin $\langle S^z\rangle(b_\perp)$ in Eq.~(\ref{eq:spin}) (red band) and the kinetic OAM $\langle L^z\rangle(b_\perp)$ in Eq.~(\ref{eq:AM}) (blue band) contributions; 
(b) the kinetic TAM density $\langle J^z\rangle(b_\perp)$ (black band) resulting from the sum of the naive TAM density $\langle J^z_{\rm naive}\rangle(b_\perp)$ in Eq.~(\ref{naive}) (blue band) and the corresponding correction term $\langle J^z_{\rm corr}\rangle(b_\perp)$ in Eq.~(\ref{corrb}) (red band);
(c) the kinetic TAM density $\langle J^z\rangle (b_\perp)$ (black band) expressed as the sum of the Belinfante-improved TAM $\langle J^z_{\rm Bel}\rangle (b_\perp)$ density in Eq.~(\ref{eq:AM_Beli}) (blue band) and the total divergence term $\langle M^z\rangle (b_\perp)$ in Eq.~(\ref{eq:M2}) (red band); 
(d) the Belinfante-improved TAM $\langle J^z_{\rm Bel}\rangle (b_\perp)$ density (black band) decomposed into its monopole $\langle J_\text{Bel}^\text{mono}\rangle (b_\perp)$ in Eq.~(\ref{lbur}) (blue band) and quadrupole $\langle J_\text{Bel}^\text{quad}\rangle (b_\perp)$ in Eq.~(\ref{lquadrup}) (red band) contributions. The bands reflect our $\alpha_s$ uncertainty of $10\%$.}
\label{Fig:totalOAM}
\end{figure}
\begin{figure}
\begin{tabular}{cc}
\subfloat[]{\includegraphics[scale=0.38]{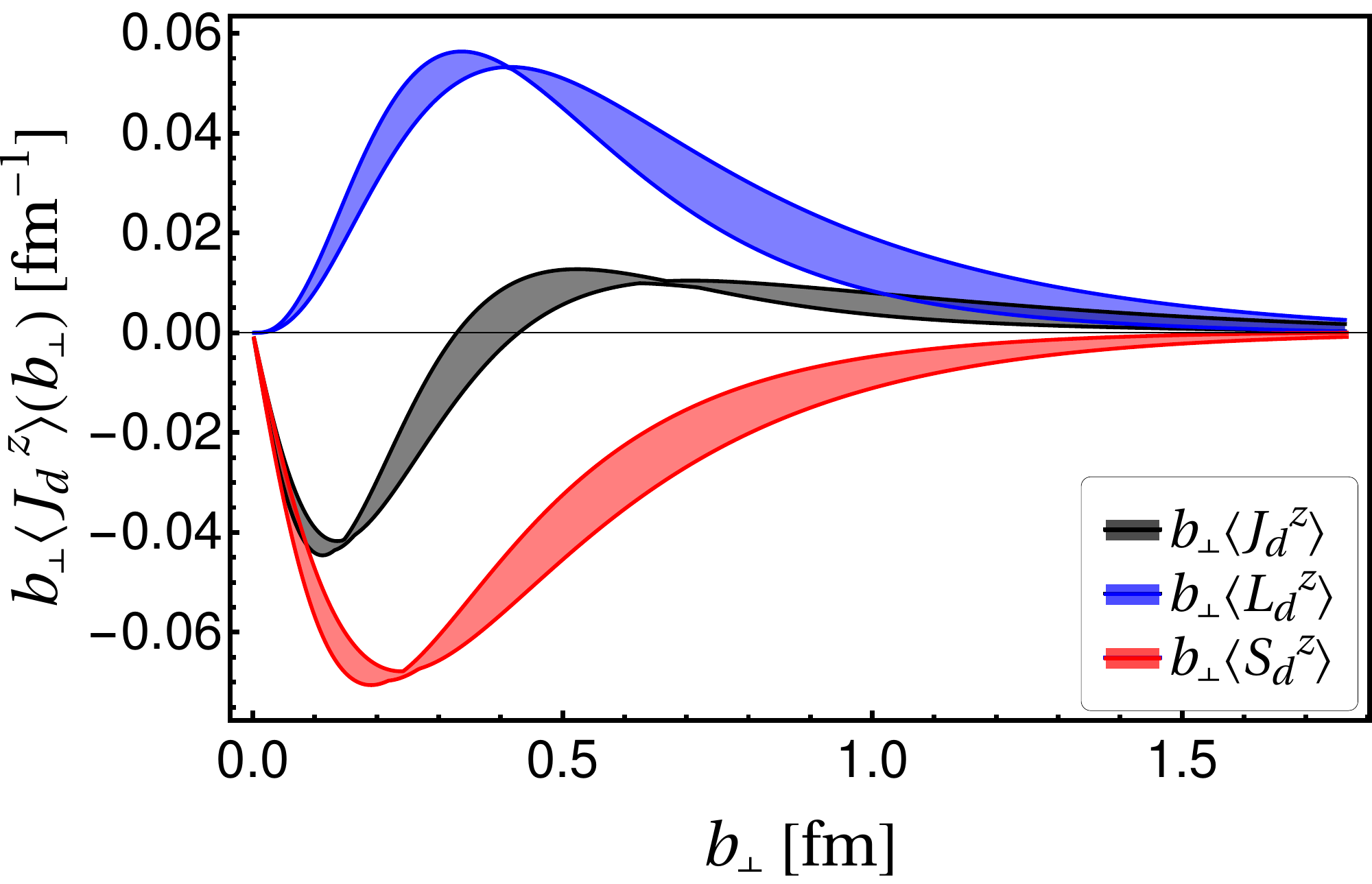}}
\end{tabular}
\begin{tabular}{cc}
\subfloat[]{\includegraphics[scale=0.38]{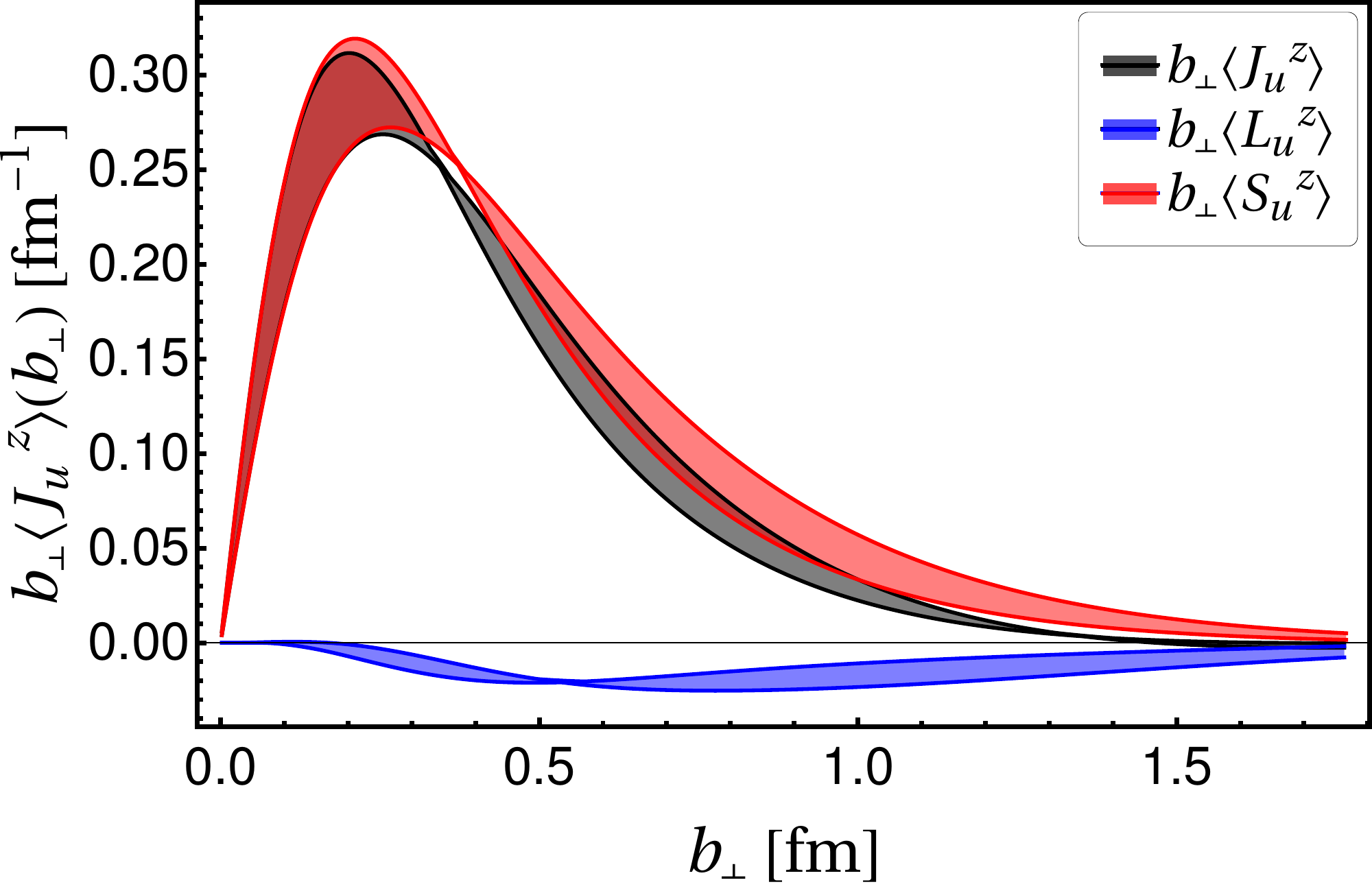}}
\end{tabular}
\begin{tabular}{cc}
\subfloat[]{\includegraphics[scale=0.38]{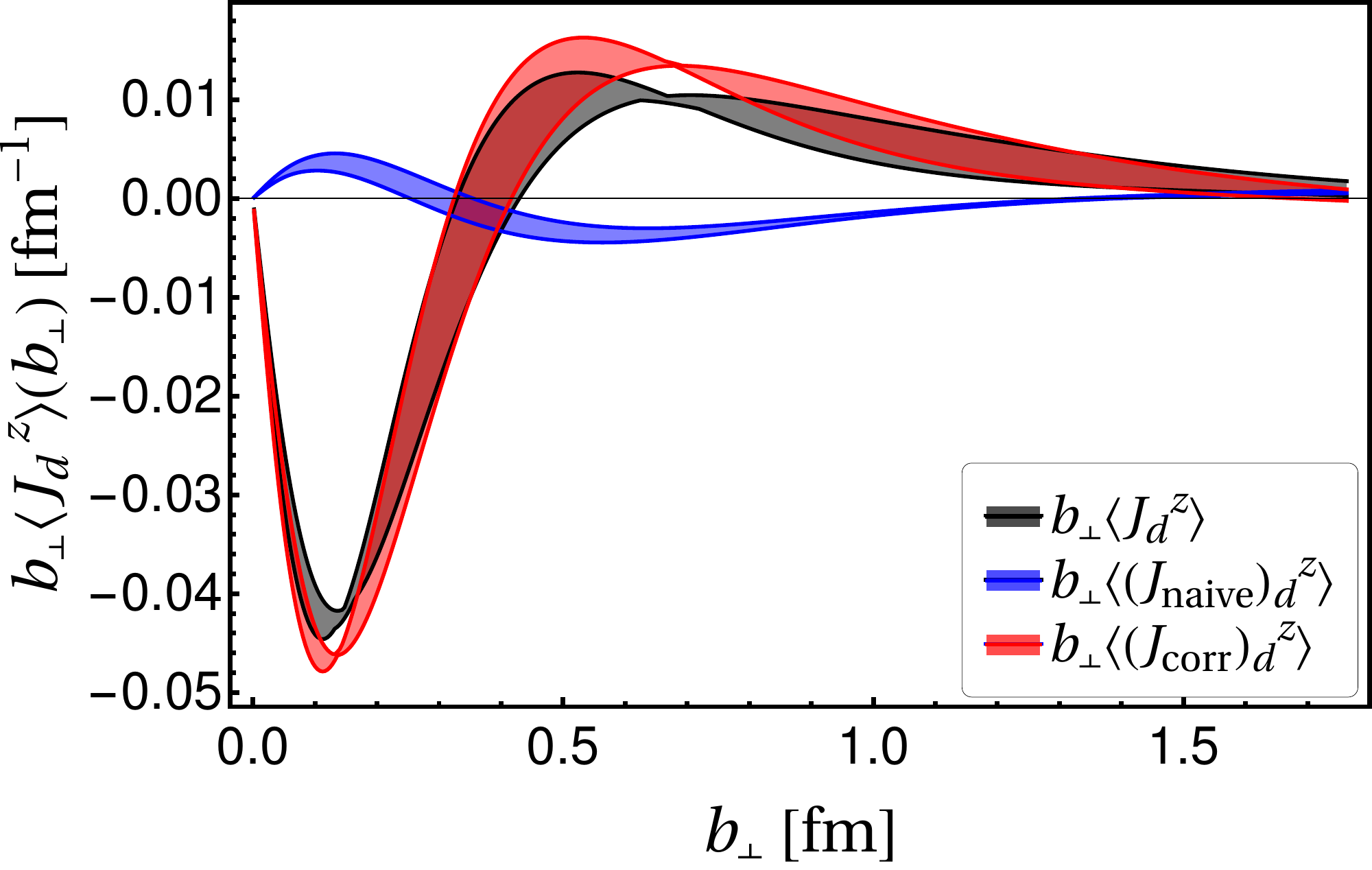}}
\end{tabular}
\begin{tabular}{cc}
\subfloat[]{\includegraphics[scale=0.38]{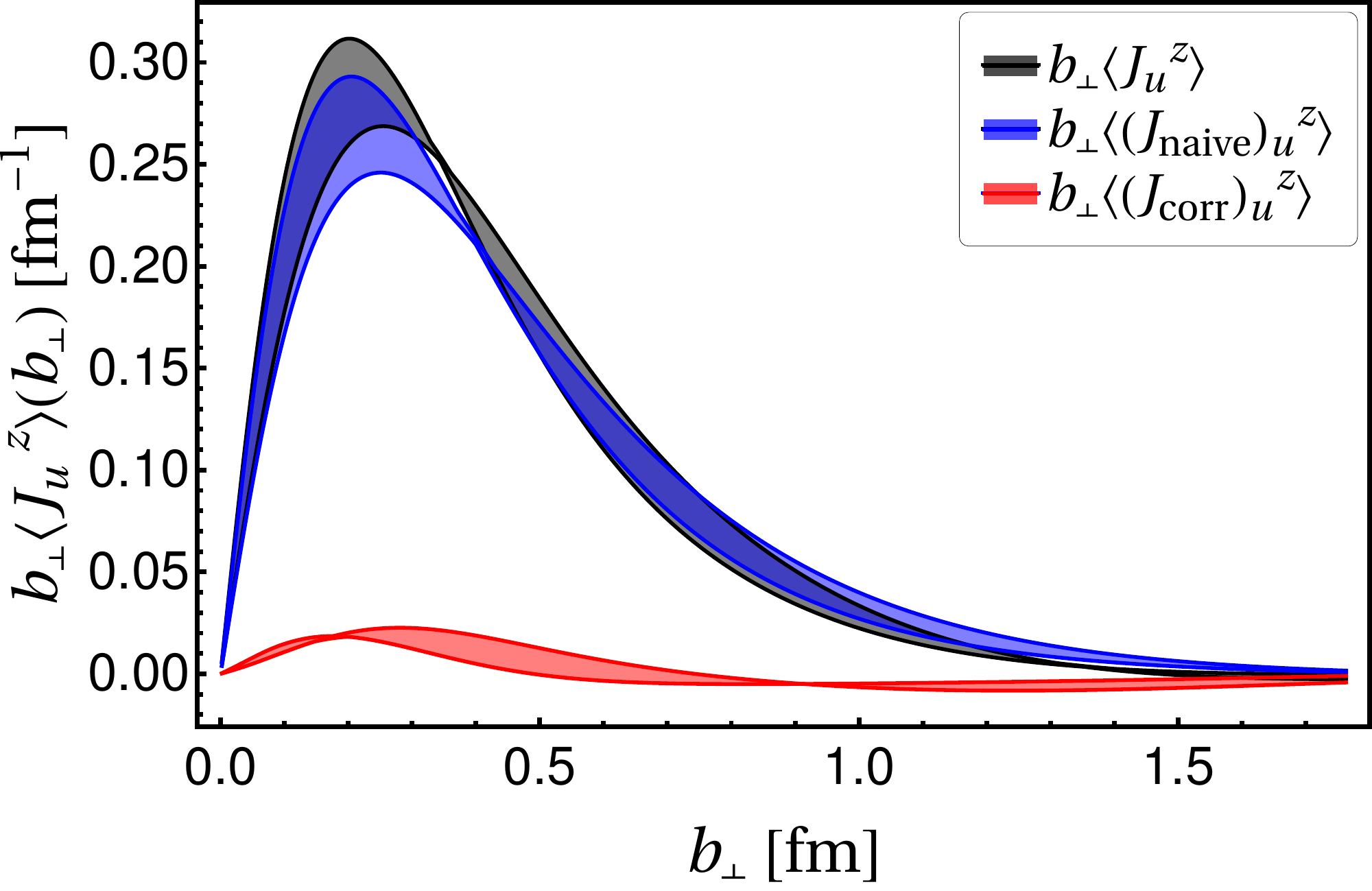}}
\end{tabular}
\begin{tabular}{cc}
\subfloat[]{\includegraphics[scale=0.38]{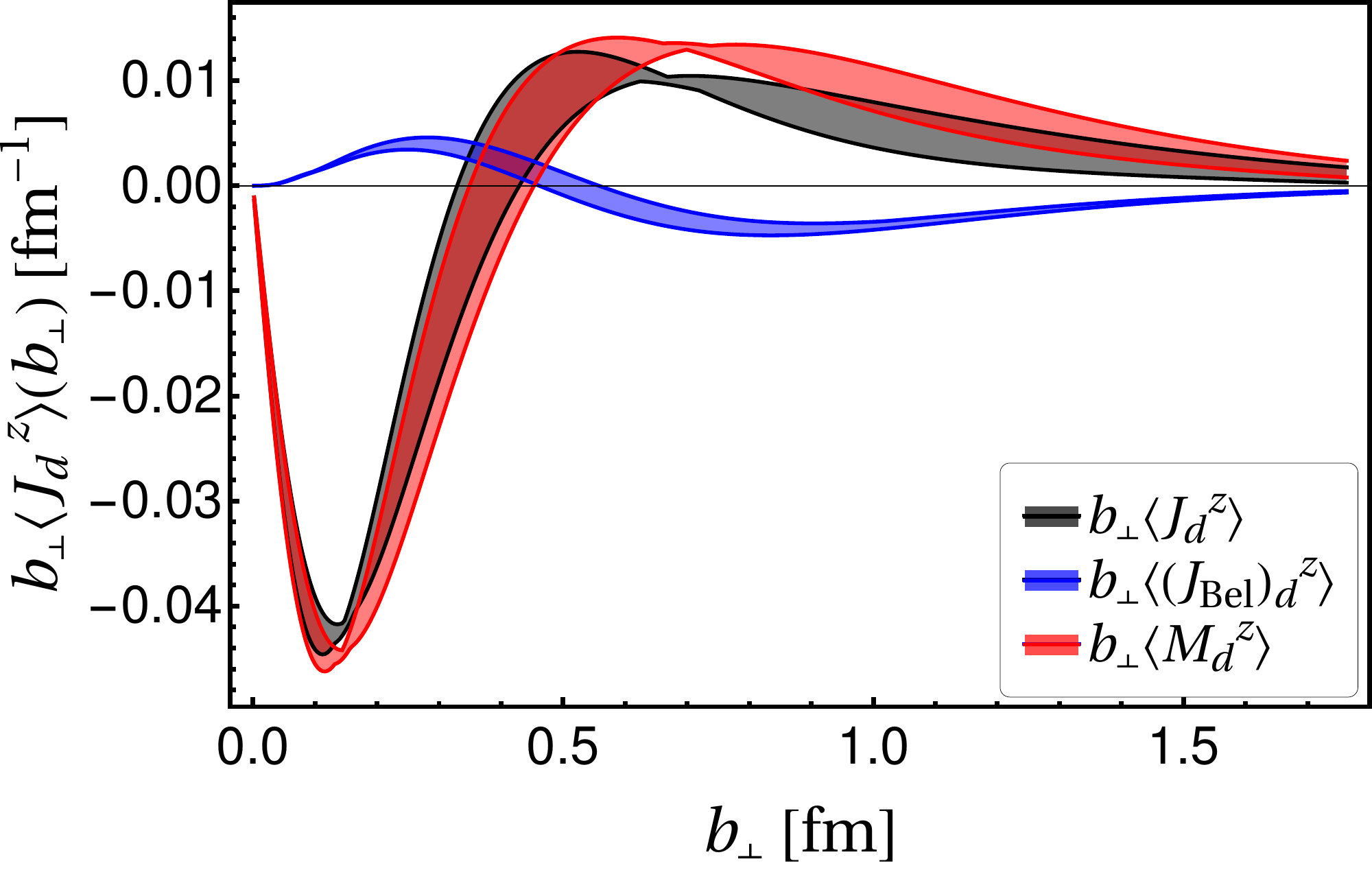}}
\end{tabular}
\begin{tabular}{cc}
\subfloat[]{\includegraphics[scale=0.38]{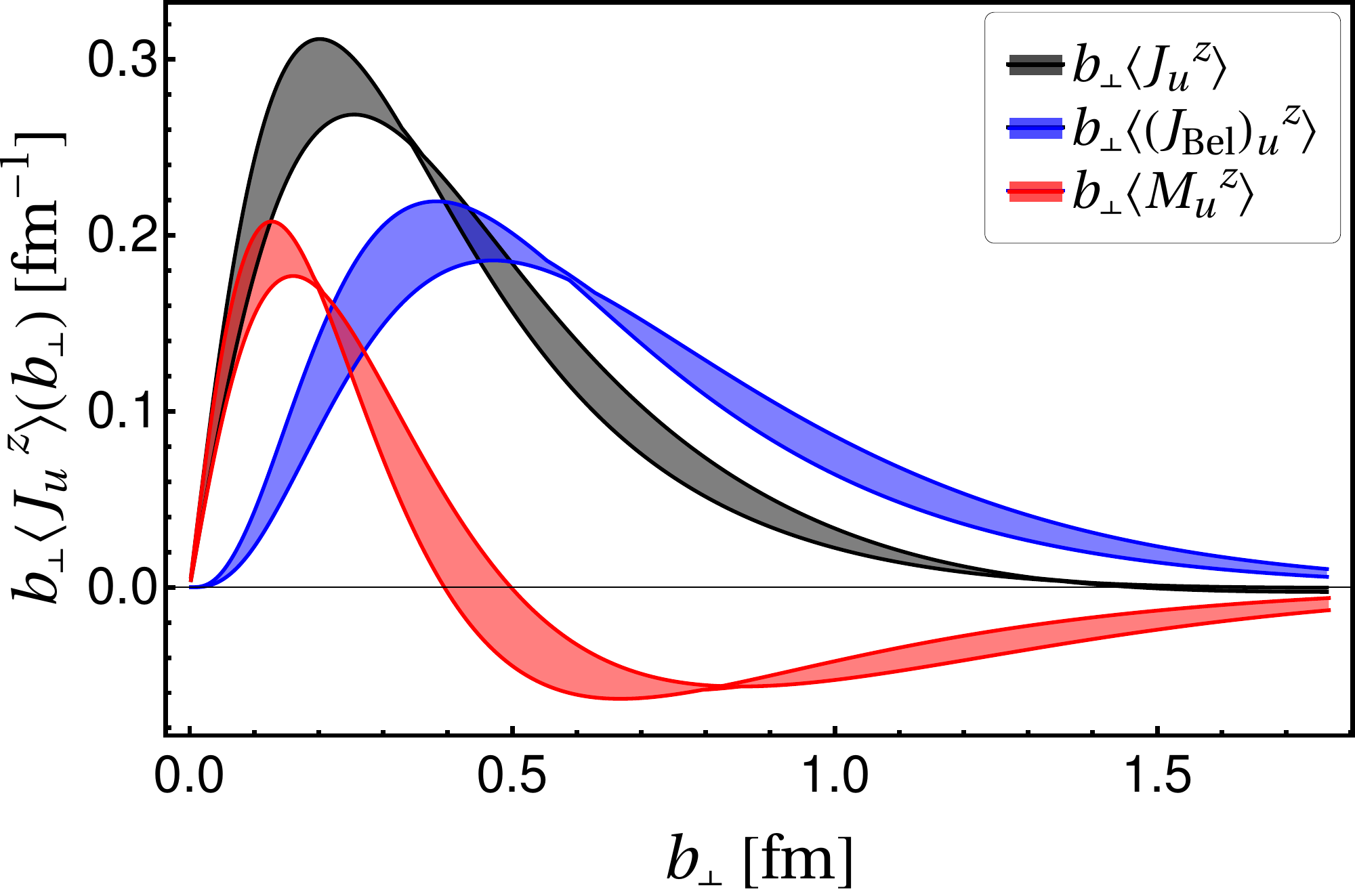}}
\end{tabular}
\begin{tabular}{cc}
\subfloat[]{\includegraphics[scale=0.38]{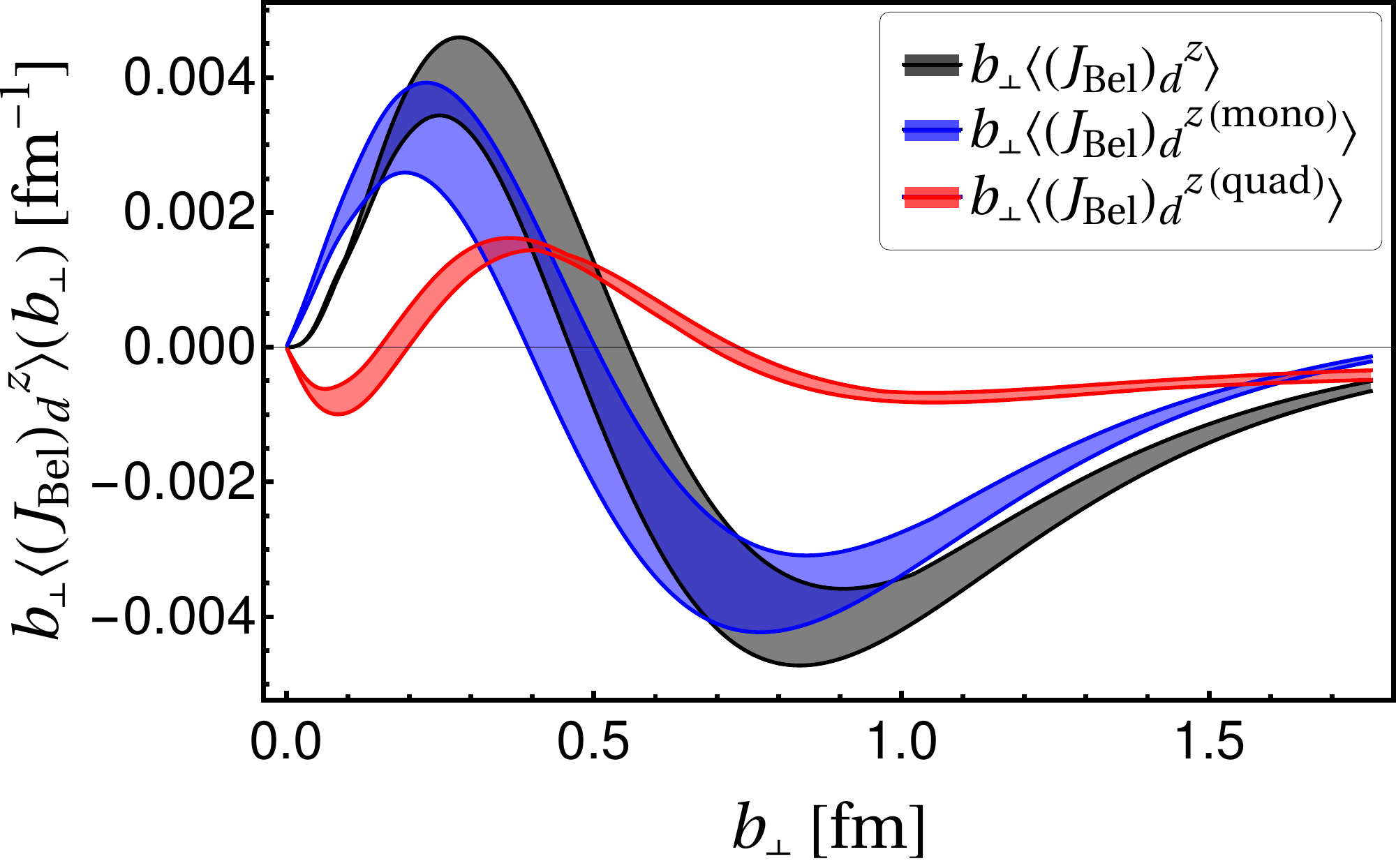}}
\end{tabular}
\begin{tabular}{cc}
\subfloat[]{\includegraphics[scale=0.38]{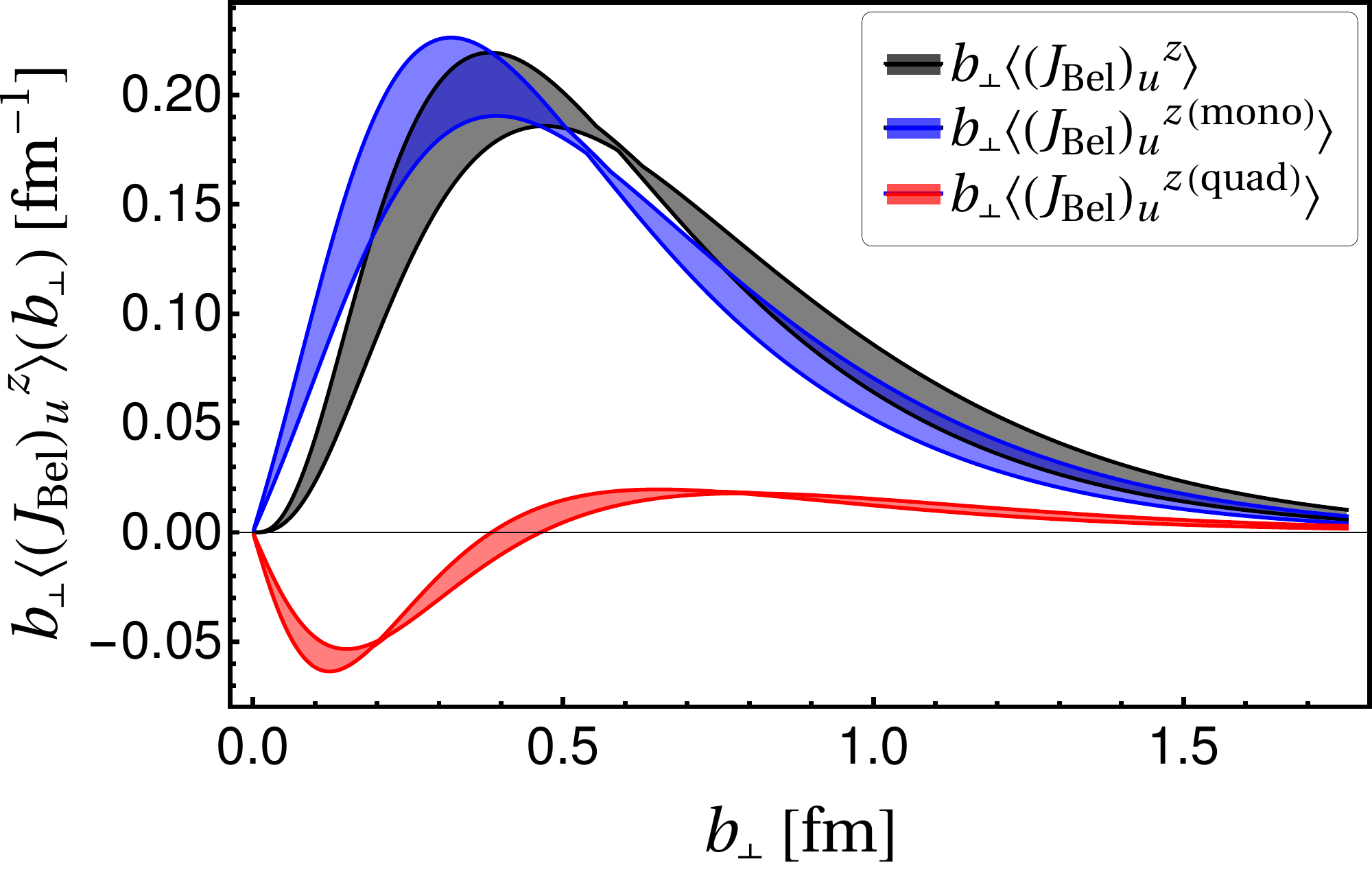}}
\end{tabular}
	\caption{Angular momentum densities of the up and the down quarks multiplied by $b_\perp$ as functions of $b_\perp$. The left panels i.e., $\{$(a), (c), (e), (g)$\}$ are for the down quark, while the right panels i.e., $\{$(b), (d), (f), (h)$\}$ are for the up quark. The legends in $\{$(a), (c), (e), (g)$\}$ or $\{$(b), (d), (f), (h)$\}$ are the same as described in $\{$(a), (b), (c), (d)$\}$ of Fig.~\ref{Fig:totalOAM} but for quarks.  
}
	\label{Fig:quarkOAM}
\end{figure}

We illustrate the valence quark GPDs of the proton in the transverse impact parameter space for zero skewness in Fig.~\ref{Fig:impact_gpds}. We observe that, except for the fact that the magnitude of $H(x,b_\perp)$ for the up quark is larger than that for the down quark, the overall nature of this distribution is the same for both the quarks. After integrating over $b_\perp$, $H(x,b_\perp)$ reduces to the ordinary unpolarized PDF $f_1(x)$ and satisfies the quark counting rule when we further integrate over $x$. Meanwhile, $E(x,b_\perp)$ for the up quark emerges as a positive distribution, whereas it is negative for the down quark. After integrating $E(x,b_\perp)$ over $x$ and $b_\perp$, we obtain the following values for the quark
anomalous magnetic moments: $\kappa_u=1.481\pm 0.029$ and $\kappa_d=-1.367\pm 0.025$ corresponding to the nucleon anomalous magnetic moments: $\kappa_p=\frac{2}{3}\kappa_u-\frac{1}{3}\kappa_d=1.443\pm 0.027$ and $\kappa_n=-\frac{1}{3}\kappa_u+\frac{2}{3}\kappa_d=-1.405\pm 0.026$, which are close to the recent results from lattice simulations: $\kappa_p^{\rm lat}=1.43(9)$ and $\kappa_n^{\rm lat}=-1.54(6)$~\cite{Alexandrou:2018sjm}. The experimental values are $\kappa_p^{\rm exp}=1.793$ and $\kappa_n^{\rm exp}=-1.913$. 

We also observe in Fig.~\ref{Fig:impact_gpds} that $E(x,b_\perp)$ falls faster than $H(x,b_\perp)$ at $x\to 1$ similar to what is found in other phenomenological models~\cite{Boffi:2007yc,Chakrabarti:2013gra,Mondal:2015uha,Maji:2017ill}. The qualitative behavior of $\widetilde{H}(x,b_\perp)$ in our approach is almost the same as in ${H}(x,b_\perp)$ but the magnitude of $\widetilde{H}(x,b_\perp)$ is relatively lower than that for ${H}(x,b_\perp)$. Meanwhile, $\widetilde{H}(x,b_\perp)$ for the down quark has the opposite sign to the up quark distribution.
 Integrating $\widetilde{H}(x,b_\perp)$ over $b_\perp$, we obtain the helicity distribution $g_1(x)$, which in our BLFQ approach is consistent with the experimental data~\cite{Xu:2021wwj}.

We further notice in Fig.~\ref{Fig:impact_gpds} that the width of all the GPDs in the transverse impact parameter space decrease as $x$ increases. This indicates that the distributions are more concentrated and the quarks are more localized near the
center of momentum ($b_\perp=0$) when they are carrying a higher longitudinal momentum  fraction. 
Meanwhile, the peaks of all the distributions shift toward to lower values of $x$ when $b_\perp$ increases. This characteristic of the distributions in the $b_\perp$-space is reassuring since the GPDs in the momentum space become broader in $-t$ with increasing $x$, as can be seen from Fig.~\ref{Fig:gpds}. On the light-front, this can be understood as the larger the momentum fraction, the smaller the kinetic energy carried by the quarks. As the total kinetic energy remains limited, the distribution in the transverse momentum broadens at higher $x$ reflecting the trend to  carry a larger portion of the kinetic energy.  As a consequence, these general features should be nearly model-independent properties of the GPDs and, indeed, they are also observed in other theoretical studies of the GPDs~\cite{Burkardt:2002hr,Chakrabarti:2013gra,Vega:2010ns,Mondal:2015uha,Chakrabarti:2015ama,Mondal:2017wbf,Maji:2017ill}.

We employ these GPDs to compute the angular momentum distributions defined in  Eqs.~(\ref{lip})-(\ref{corrb}) following Eqs.~(\ref{eq:Lb})-(\ref{eq:Gb}). In Fig.~\ref{Fig:totalOAM}, we illustrate different definitions of the TAM densities summing over the up and the down quark contributions. We display the distribution $b_\perp \langle J^z\rangle(b_\perp)$ as a function of $b_\perp$ and notice that the TAM is positive over all $b_\perp$. It has the peak near $b_\perp\sim 0.3$ fm, falls slowly with increasing $b_\perp$ and becomes very small near $b_\perp\sim 1.5$ fm. The error bands in our distributions are due to the $10\%$ uncertainties in the coupling constant. 

Figure~\ref{Fig:totalOAM}(a) shows the kinetic TAM density $\langle J^z\rangle (b_\perp)$  
as the sum of the spin $\langle S^z\rangle(b_\perp)$ and the kinetic OAM $\langle L^z\rangle(b_\perp)$ contributions each multiplied by $b_\perp$. Both the contributions show positive distributions. 
In contrast to the results in a quark-diquark model~\cite{Lorce:2017wkb}, where the OAM component is larger than the spin component of the TAM density, the spin distribution in our approach strongly dominates over the kinetic OAM density. 
The $\langle L^z\rangle (b_\perp)$ is mainly effective over the range $0.2<b_\perp<0.6$ fm. 
It should be noted that these results are obtained within the valence Fock representation, while the higher Fock components $|qqqg\rangle$ and $|qqqq\bar{q}\rangle$ are anticipated to have significant effects on the spin and OAM distributions. With the inclusion of dynamical gluons and sea quarks, the quark spin contribution may be suppressed, and the OAM can play an enhanced role in the TAM density.

Figure~\ref{Fig:totalOAM}(b) compares the kinetic TAM $\langle J^z\rangle(b_\perp)$ and the naive distribution $\langle J^z_{\rm naive}\rangle(b_\perp)$. Their difference, attributed to the correction term $\langle J^z_{\text{corr}\rangle}(b_\perp)$ in Eq.~(\ref{corrb}), is also shown in this plot. We find that $\langle J^z_{\rm naive}\rangle(b_\perp)$ is close to $\langle J^z\rangle(b_\perp)$. The $\langle J^z_{\text{corr}\rangle}(b_\perp)$ exhibits a negative central region  surrounded by a ring of positive distribution, which is in accord with the behavior observed in the quark-diquark model~\cite{Lorce:2017wkb}. 

We present the comparison between the kinetic TAM $\langle J^z\rangle(b_\perp)$ and the Belinfante-improved density $\langle J^z_\text{Bel}\rangle(b_\perp)$ in Fig.~\ref{Fig:totalOAM}(c). In the same plot, we also show their difference given by the total divergence $\langle M^z\rangle(b_\perp)$ term in Eq.~(\ref{divb}). The $\langle J^z_\text{Bel}\rangle(b_\perp)$ is smaller at $b_\perp<0.4$ fm but larger at $b_\perp>0.4$ fm than the $\langle J^z\rangle(b_\perp)$. The Belinfante-improved density falls slower at higher $b_\perp$ than the TAM density. We also find that the peak of the Belinfante-improved distribution is lower and appears at higher value of $b_\perp$ compared to that for the TAM density. Also, the Belinfante-improved distribution is broader and falls more slowly in $b_\perp$ compared with the kinetic TAM $\langle J^z\rangle(b_\perp)$. It can be noticed that the $\langle M^z\rangle(b_\perp)$
has a positive core and a negative tail. $\langle M^z\rangle(b_\perp)$ has a significant contribution to the TAM distribution and this can be ascribed from the fact that it is related to the spin distribution, which in our model provides the dominating contribution to the TAM density.

As another illustration of our results, the decomposition of the Belinfante-improved TAM 
 in term of its monopole $\langle J_\text{Bel}^{z(\text{mono})}\rangle (b_\perp)$ and quadrupole $\langle J_\text{Bel}^{z(\text{quad})}\rangle (b_\perp)$ contributions is presented in Fig.~\ref{Fig:totalOAM}(d). One notices that the qualitative behavior of $\langle J_\text{Bel}{z(\text{mono})}\rangle (b_\perp)$ and $\langle J_\text{Bel}^{z(\text{quad})}\rangle (b_\perp)$ distributions is similar to $\langle J^z_{\rm naive}\rangle(b_\perp)$ and $\langle J^z_{\text{corr}}\rangle(b_\perp)$, respectively. Finally, we observe that the correction term $\langle J^z_\text{corr}\rangle(b_\perp)$, the total divergence term $\langle M^z\rangle(b_\perp)$, and the quadrupole contribution $\langle J^{z(\text{quad})}_\text{Bel}\rangle(b_\perp)$ integrate to zero. However, at the density level, we need to take them into account while comparing different definitions for the angular momentum distribution. Note that these findings in our BLFQ approach are also supported by the analysis based on a light-front quark-diquak model~\cite{Lorce:2017wkb}. 
 
In Fig.~\ref{Fig:quarkOAM}, we demonstrate the angular momentum densities for quark  flavors by considering all the different definitions described above.  In Fig.~\ref{Fig:quarkOAM}(a) and \ref{Fig:quarkOAM}(b), we present the kinetic TAM $\langle J^z\rangle^q(b_\perp)=\langle S^z\rangle^q(b_\perp)+\langle L^z\rangle^q(b_\perp)$ for the up and the down quarks, respectively  multiplied by $b_\perp$. In our BLFQ approach, the contribution in $\langle J^z\rangle(b_\perp)$ from $\langle S^z\rangle(b_\perp)$ is larger than that from $\langle L^z\rangle(b_\perp)$ for the up quark, whereas for the down quark, $\langle S^z\rangle(b_\perp)$ dominates at lower $b_\perp$ but $\langle L^z\rangle(b_\perp)$ is superior at large distance. For the up quark, the spin and the OAM densities show positive and negative distributions, respectively, while they are opposite for the down quark. In essence, $\langle J^z\rangle(b_\perp)$ for the down quark exhibits a negative core near the center of momentum of the proton and it has a positive tail at large distance. Meanwhile, $\langle J^z\rangle(b_\perp)$ is almost equivalent to $\langle S^z\rangle(b_\perp)$ for the up quark and they show positive distributions. 

In Fig.~\ref{Fig:quarkOAM}(c) and Fig.~\ref{Fig:quarkOAM}(d), we compare the up and the down quark kinetic TAM $\langle J^z\rangle(b_\perp)$ with their naive density $\langle J^z_{\rm naive}\rangle(b_\perp)$, respectively. The naive density $\langle J^z_{\rm naive}\rangle(b_\perp)$ is positive for the up quark, while for the down quark, it is also positive at low $b_\perp$ but negative at higher $b_\perp$. The difference between them is provided by the correction term $ \langle J^z_{\text{corr}}\rangle(b_\perp)$ in Eq.~\eqref{corrb}. The correction term for the up quark is small but for the down quark it is large and close to $\langle J^z\rangle(b_\perp)$. Meanwhile, the comparison between the kinetic TAM $\langle J^z\rangle(b_\perp)$ and the Belinfante-improved TAM $\langle J^z_\text{Bel}\rangle(b_\perp)$ for the up and the down quark is illustrated in Fig.~\ref{Fig:quarkOAM}(e) and Fig.~\ref{Fig:quarkOAM}(f), respectively. The difference is described by the $ \langle M^z\rangle (b_\perp)$ term in Eq.~\eqref{divb}. For the down quark, the $\langle J^z_\text{Bel}\rangle(b_\perp)$ is very small compared to $\langle J^z\rangle(b_\perp)$   and the major contribution in the kinetic TAM is coming from $ M^z(b_\perp)$. However, they are comparable for the up quark. In our BLFQ approach the qualitative behavior of the naive and the Belinfante-improved densities is very similar for the down quark but different for the up quark.

Finally, the monopole and quadrupole contributions to the Belinfante-improved total density for the quarks are shown in Fig.~\ref{Fig:quarkOAM}(g) and \ref{Fig:quarkOAM}(h). The monopole contribution dominates over the quadrupole contribution for both the quarks. Note that the integrations of the correction term $\langle J^z_\text{corr}\rangle(b_\perp)$, the total divergence term $\langle M^z\rangle(b_\perp)$, and the quadrupole contribution $\langle J^{z(\text{quad})}_\text{Bel}\rangle(b_\perp)$ for the individual quarks are also zero but all the terms need to be retained while comparing results among different definitions at the level of  distributions.

Summing over the flavors, we have obtained the total spin contributed by the quarks to
the proton spin. Within our model that incorporates only the valence Fock sector, we have found that the quark spin at the model scale contributes $\sim 91\%$ to the proton spin, whereas the contribution of quark spin captures only $\sim 40\%$ as revealed from the experiment~\cite{Leader:2010rb}. This evident discrepancy suggests the need to extend our model to append the higher Fock sectors, which have a significant effects on the proton spin. With dynamical gluons and sea quarks, the quark spin contribution can be reduced and the OAM can play a substantial role in understanding the nucleon spin. Simultaneously, the gluon and sea quark contributions to the total spin will emerge. Meanwhile, the QCD scale evolution also needs to be  taken into account.

\section{SUMMARY}\label{sec:summary}
Using a recently proposed light-front model for the proton based on a Hamiltonian formalism, we studied its valence quark GPDs. The effective Hamiltonian incorporates light-front holography, longitudinal confinement, and the one gluon exchange interaction for the valence quarks suitable for low-resolution properties. We obtained the proton LFWFs as the eigenvectors of this Hamiltonian by solving its mass eigenstates using BLFQ as a relativistic three-quark problem. The parameters in this BLFQ model have previously  been adjusted by fitting the nucleon mass and the flavor Dirac form factors. We then employed the LFWFs to compute the valence quark unpolarized and helicity dependent GPDs of the proton. We presented  results for the GPDs  in both momentum space and position space for zero skewness and we found that the qualitative behavior of the GPDs in our BLFQ approach bears similarities to other phenomenological models. 

We have  employed these GPDs to study the various definitions of angular momentum at the density level. Within our model that includes only the leading Fock sector, we have found that the spin contribution to the TAM strongly dominates over the OAM distribution in contrast to the results in a quark-diquark model~\cite{Lorce:2017wkb}. On the other hand, the naive angular momentum density is found to be close to the TAM distribution. Meanwhile, we have observed that the Belinfante-improved angular momentum density is distinctly different from the TAM distributions, whereas the difference between them given by the total divergence term has a positive core surrounded by a negative tail. When we decomposed the Belinfante-improved angular momentum density into its monopole and quadrupole components, we noticed that the dominating contribution comes from the monopole density. In our approach, we have illustrated explicitly that no discrepancies were found between different definitions of angular momentum when all the terms integrating to zero are included in the expressions. These findings in our BLFQ approach are consistent with the analysis based on a light-front quark-diquak model~\cite{Lorce:2017wkb}.

We have subsequently presented the up and the down quarks' TAM densities and found that the up quark contribution to the TAM distribution is much larger than that for the down quark. The spin and OAM densities are comparable but opposite in sign for the down quark, while for the up quark, the spin distribution is much stronger than the OAM density. The naive distributions for both the quarks were found to be similar to  their TAM densities. Although the Belinfante-improved and TAM distributions are similar for the down quark, they are different from each other for the up quark in our current approach.

\section*{ACKNOWLEDGMENTS}
C. M. is supported by new faculty start up funding by the Institute of Modern Physics, Chinese Academy of Sciences, Grant No. E129952YR0. 
C. M. also thanks the Chinese Academy of Sciences Presidents International Fellowship Initiative for the support via Grants No. 2021PM0023.  X. Z. is supported by new faculty startup funding by the Institute of Modern Physics, Chinese Academy of Sciences, by Key Research Program of Frontier Sciences, Chinese Academy of Sciences, Grant No. ZDB-SLY-7020, by the Natural Science Foundation of Gansu Province, China, Grant No. 20JR10RA067 and by the Strategic Priority Research Program of the Chinese Academy of Sciences, Grant No. XDB34000000. J. P. V. is supported by the Department of Energy under Grants No. DE-FG02-87ER40371, and No. DE-SC0018223 (SciDAC4/NUCLEI). This research used resources of the National Energy Research Scientific Computing Center (NERSC), a U.S. Department of Energy Office of Science User Facility operated under Contract No. DE-AC02-05CH11231.
 A portion of the computational resources were also provided by Gansu Computing Center.

\end{document}